\documentclass[aps,prb,reprint,twocolumn,superscriptaddress,nobibnotes,nofootinbib,floatfix]{revtex4-2}
\usepackage{amsmath,amssymb}
\usepackage{bm}
\usepackage{graphicx}
\usepackage{xcolor}
\usepackage{booktabs}
\usepackage[hidelinks]{hyperref}
\makeatletter
\def\frontmatter@thefootnote{%
  \ifcase\csname c@\@mpfn\endcsname
  \or\textdagger\or\textasteriskcentered
  \else\@fnsymbol{\csname c@\@mpfn\endcsname}\fi}
\makeatother
\newcommand{\tr}{\operatorname{tr}}
\newcommand{\ii}{\mathrm{i}}
\newcommand{\ee}{\mathrm{e}}

\begin{document}
\title{Quantized response of amplification in non-Hermitian photonic time quasicrystals}
\author{Shuhang Chen}
\thanks{These authors contributed equally to this work.}
\affiliation{Department of Electronic Engineering and Information Science, University of Science and Technology of China, Hefei, Anhui 230027, China}
\author{Zhi Zheng}
\thanks{These authors contributed equally to this work.}
\affiliation{CAS Key Laboratory of Strongly-Coupled Quantum Matter Physics, and Department of Physics, University of Science and Technology of China, Hefei, Anhui 230026, China}
\affiliation{International Center for Quantum Design of Functional Materials (ICQD), Hefei National Research Center for Physical Sciences at the Microscale, University of Science and Technology of China, Hefei, Anhui 230026, China}
\affiliation{Hefei National Laboratory, University of Science and Technology of China, Hefei, Anhui 230088, China}
\author{Qi Zhu}
\email{zhuqi@ustc.edu.cn}
\affiliation{Department of Electronic Engineering and Information Science, University of Science and Technology of China, Hefei, Anhui 230027, China}
\begin{abstract}
Photonic time crystals amplify waves through periodic temporal modulation of the permittivity. We study a non-Hermitian photonic time quasicrystal, a temporal analogue of the non-Hermitian Aubry--Andr\'e--Harper model of quasiperiodic lattices. Its modulation is incommensurate and has a complex phase, whose imaginary part, the complexification parameter, acts as an imaginary gauge field on the modulation harmonics. We find that the bulk amplification rate, as a function of this parameter, is organized into distinct phases, within each of which its slope is locked to an integer, known in quasiperiodic operator theory as Avila's acceleration. The phase boundaries are predicted from Fourier coefficients of rational approximants and confirmed by transfer-matrix and time-domain simulations. Under analytic distortions of the modulation waveform, the integer phases persist while their boundaries move. Consequently, two input frequencies in different integer phases have an amplification ratio that varies exponentially with the complexification parameter, which enables spectral selection. These results establish a topologically protected, quantized response law for amplification in such time-varying media.
\end{abstract}
\maketitle

\begin{figure*}[t]
\centering
\includegraphics[width=\textwidth]{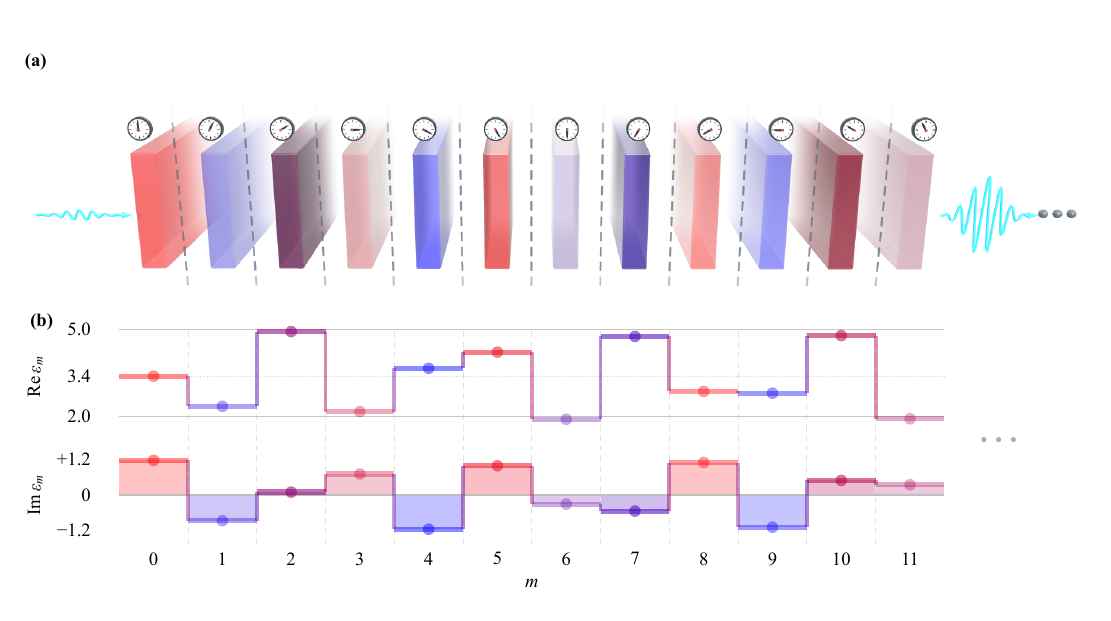}
\caption{\textbf{A non-Hermitian photonic time quasicrystal.}
(a) A wave propagating through a spatially uniform medium with quasiperiodically switched permittivity. Greater block opacity denotes a larger real permittivity; the color runs from blue for negative to red for positive imaginary permittivity. Clock faces mark the time at each temporal layer, advancing by equal steps because the layers have equal duration; the enlarged outgoing waveform illustrates amplification.
(b) Real and imaginary parts of the complex permittivity for the displayed segment at $h=1.0$ and $\varphi_0=0$. The imaginary component changes sign and has zero average over the modulation phase.}
\label{fig:one}
\end{figure*}

Modulating material parameters in time breaks time-translation symmetry and makes time a new dimension for controlling light waves~\cite{Yuan2018,Galiffi2022,Pendry2022,Engheta2023}. The paradigmatic case is the photonic time crystal (PTC), a spatially uniform medium whose permittivity varies periodically in time and opens momentum gaps, inside which waves grow exponentially~\cite{Zurita2009,ReyesAyona2015,Lyubarov2022}. These momentum gaps are also topological, carrying an invariant tied to the relative phase of the forward- and backward-propagating waves and supporting edge states that live in time~\cite{Lustig2018}. PTCs have now been extended to non-Hermitian modulation~\cite{Wang2018,KoutserimpasAlu2018,LiYinGaliffiAlu2021} and free-electron radiation~\cite{Dikopoltsev2022}, observed in plasmonic metamaterials~\cite{Guo2026}, and carried into quantum electrodynamics~\cite{Bae2026}.

Periodic driving, however, is only one way to organize temporal modulation. Under random modulation, the drift of a pulse can be suppressed while its amplitude keeps growing, in contrast with spatial Anderson localization~\cite{Sharabi2021}. Incommensurate modulation instead produces a photonic time quasicrystal (PTQC), which retains momentum gaps despite the absence of temporal periodicity~\cite{Aperiodic2025,Fibonacci2025}. Its richer gap structure is classified topologically by Chern numbers defined over the Floquet and modulation phases~\cite{Ni2025}, as in spatial quasicrystals~\cite{Kraus2012,Verbin2013,KrausZilberberg2016}. These gaps also amplify waves~\cite{Aperiodic2025,Fibonacci2025,Ni2025}, at a rate that depends on the gap and the input frequency. These developments raise the question of whether topology can constrain amplification itself, beyond classifying the gaps in which it occurs.

A precedent exists in spatial non-Hermitian quasicrystals, where topology constrains a rate, the rate at which eigenstates decay in space. Spectral winding around a point gap~\cite{Bergholtz2021} is tied to this decay rate~\cite{Longhi2019a,Longhi2019b,Zeng2020,Jiang2019,Cai2021}, a link observed as a localization transition in photonic quantum walks~\cite{Weidemann2022,Lin2022}. When the modulation phase is analytically continued into the complex plane, Avila's global theory identifies the integer behind this link~\cite{Liu2021,Avila2015,AvilaJitomirskayaSadel2014,WangIMRN2024,HanSchlag}. Along this continuation, the Lyapunov exponent that gives the decay rate varies linearly with the imaginary part of the phase, and its slope is an integer that counts a winding in the modulation phase. This exponent, however, is an inverse localization length, so the integer has to be extracted indirectly from localization lengths at different imaginary parts of the phase. In a photonic time quasicrystal, the modulation runs in time, and the same exponent becomes the rate at which a wave grows. Whether the integer can thereby appear directly as a response of amplification remains unexplored.

Here we analytically continue the phase of a quasiperiodic permittivity modulation into the complex plane, which turns a Hermitian photonic time quasicrystal into a non-Hermitian one. We find that the slope of the amplification rate with respect to the imaginary part of the phase is precisely this integer, Avila's acceleration. Even with a single harmonic in the modulation, the integer climbs a staircase of successive values, whereas in the spatial non-Hermitian Aubry--Andr\'e--Harper (AAH) model it changes at most once~\cite{Longhi2019a,Liu2021,Avila2015}. We turn the rational approximants used in Avila's proof into an explicit method that predicts the transitions from two Fourier coefficients at a single value of the imaginary part, as confirmed by transfer-matrix calculations and full-wave time-domain simulations. The integers persist under analytic distortions of the modulation waveform, which only move the phase boundaries. For a wave with two frequency components, the difference of their integers sets how fast the output power ratio changes, so that one component is amplified selectively over the other. Non-Hermiticity introduced through a complex phase thereby becomes a topological dial for amplification.

\begin{figure*}[t]
\centering
\includegraphics[width=\textwidth]{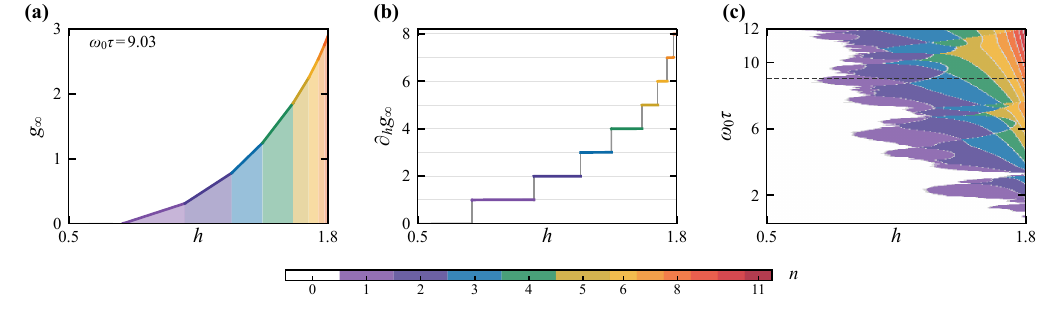}
\caption{\textbf{Quantized growth response and integer phase diagram.}
(a) Bulk growth rate $g_\infty$ at $\omega_0\tau=9.03$ and
(b) its slope $\partial_h g_\infty$ with respect to the complexification parameter $h$.
(c) Integer phases in the $(h,\omega_0\tau)$ plane; color encodes the integer $n$, with white for $n=0$ and for unclassified points near phase boundaries (Supplemental Material, Sec.~S6\,A), and the horizontal line marks $\omega_0\tau=9.03$, the frequency used in (a,b).}
\label{fig:two}
\end{figure*}

We start from a spatially uniform dielectric with constant permeability
$\mu_0$, whose real permittivity is switched between temporal layers of
equal duration $\tau$. Following the photonic time-quasicrystal framework~\cite{Ni2025}, we take
$\varepsilon_m=\varepsilon_R+\delta\varepsilon\sin X_m$, where
$m=0,1,2,\ldots$ labels the temporal layers,
$X_m=2\pi(\alpha m+\varphi_0)$ with irrational $\alpha$, and $\varphi_0$
is the initial modulation phase. Analytically continuing the modulation phase $X_m$ into the complex plane
gives the non-Hermitian PTQC illustrated in Fig.~\ref{fig:one}(a),
\begin{equation}
\varepsilon_m=\varepsilon_R+\delta\varepsilon\sin(X_m+\ii h).
\label{eq:eps}
\end{equation}
Here $h$, the imaginary part of the modulation phase, is the complexification parameter; $h=0$ recovers the Hermitian PTQC~\cite{Ni2025}.

This permittivity sequence has the same complexified single-harmonic form as the on-site potential of the spatial non-Hermitian AAH model~\cite{Longhi2019a,Longhi2019b}. Here, however, Maxwell's equations give a wave equation whose coefficient is proportional to $1/\varepsilon_m$ [Eq.~\eqref{eq:hill}], and the wave accumulates phase over each layer of finite duration, so the layer transfer matrix differs from that of the AAH model while remaining an analytic cocycle to which Avila's theory applies.

The real and imaginary parts follow directly from this continuation,
\begin{equation}
\begin{aligned}
\operatorname{Re}\varepsilon_m&=\varepsilon_R+\delta\varepsilon\cosh h\,\sin X_m,\\
\operatorname{Im}\varepsilon_m&=\delta\varepsilon\sinh h\,\cos X_m.
\end{aligned}\label{eq:quad}
\end{equation}
Their modulation amplitudes are linked by $h$, and their phases differ by
a quarter cycle [Fig.~\ref{fig:one}(b)]. The imaginary part thus describes gain and loss that alternate along the sequence and average to zero over the modulation phase. In terms of the harmonics $e^{\pm\ii X_m}$, non-Hermiticity weights the positive- and negative-frequency components of the modulation unequally, by factors $e^{\pm h}$, whereas a real modulation contains them with equal weights. Throughout,
$\varepsilon_R=3.4$, $\delta\varepsilon=1$, and
$\alpha=(\sqrt5-1)/2$ unless stated otherwise.

Spatial uniformity lets each wavevector $k$ evolve on its own, and the displacement field $D_k(t)$ obeys a Hill equation with a piecewise-constant coefficient,
\begin{equation}
\ddot D_k+\omega_0^2\frac{\varepsilon_R}{\varepsilon_m}D_k=0,\label{eq:hill}
\end{equation}
where $\omega_0=c|k|/\sqrt{\varepsilon_R}$ is the frequency in the
unmodulated reference medium and $\omega_0\tau$ its phase accumulation
over one layer. Across a temporal interface without impulsive sources,
$D_k$ and $B_k$ remain continuous. Constant permeability then also
ensures continuity of $\dot D_k$ (Supplemental Material~\cite{SupplementalMaterial}, Sec.~S1\,A).
The electric field generally jumps when the permittivity switches.
Let $v_m=(D_k,\ii\dot D_k/\omega_0)^{\mathsf T}$ denote the state at
the beginning of layer $m$. Propagation gives $v_{m+1}=S_m v_m$, with the layer matrix
\begin{equation}
S_m=\begin{pmatrix}
\cos(\omega_0\tau z_m)&-\ii\sin(\omega_0\tau z_m)/z_m\\
-\ii z_m\sin(\omega_0\tau z_m)&\cos(\omega_0\tau z_m)
\end{pmatrix}.
\label{eq:S}
\end{equation}
Here $z_m^2=\varepsilon_R/\varepsilon_m$. All entries are even in $z_m$,
so the square root introduces no branch ambiguity. After $N$ layers,
$v_N=M_Nv_0$, where the transfer matrix is $M_N=S_{N-1}\cdots S_0$.

Each layer matrix has $\det S_m=1$, so the two temporal solutions in each layer change by reciprocal factors: when one grows, the other decays at the same rate. The modulation can still supply energy and amplify waves exponentially, as in Hermitian PTC momentum gaps~\cite{Lyubarov2022} (Supplemental Material, Secs.~S1\,A and S5\,A). The transfer matrix remains analytic in the complex phase provided the
permittivity avoids zero. For this modulation, the analytic strip is
$|h|<h_\star=\operatorname{arccosh}(\varepsilon_R/\delta\varepsilon)$,
with $h_\star\approx1.89$. We use an instantaneous, nondispersive complex
permittivity, as in earlier studies of non-Hermitian time-periodic
media~\cite{Wang2018}.

\begin{figure}[t]
\centering
\includegraphics[width=\columnwidth]{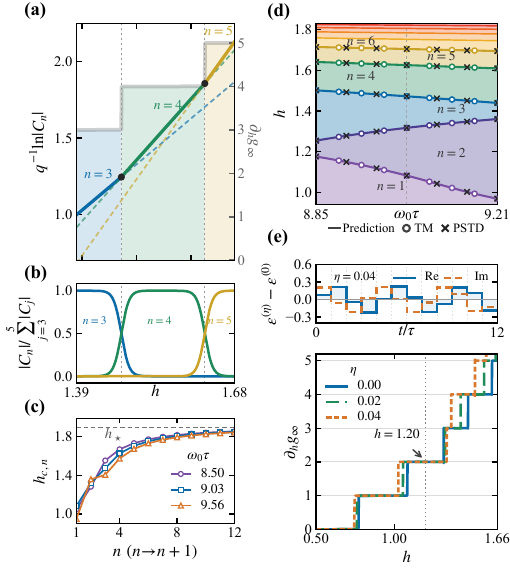}%
\caption{\textbf{Persistence and transitions of integer phases.}
(a) Fourier lines $q^{-1}\ln|C_n^{(q)}|$ for orders 3, 4, 5 at $\omega_0\tau=9.03$, $p/q=89/144$; solid segments denote dominance. Translucent filling shows the numerical response (right axis); dots mark coefficient crossings.
(b) Relative Fourier magnitudes $|C_n^{(q)}|/\sum_{j=3}^{5}|C_j^{(q)}|$, normalized over the three displayed orders to reveal the dominant order and its exchange across transitions.
(c) Predicted $n\to n+1$ boundaries at three frequencies, through $q=1597$; the dashed line marks the analytic-strip edge $h_\star$.
(d) Integer phases colored as in Fig.~\ref{fig:two}, separated by predicted boundaries through $q=987$. Transfer-matrix (TM) circles and carrier-envelope pseudospectral time-domain (PSTD) crosses trace the boundaries from $n=1\to2$ to $5\to6$; the dotted line marks $\omega_0\tau=9.03$.
(e) The distortion itself, $\varepsilon^{(\eta)}-\varepsilon^{(0)}$, at $\eta=0.04$, $h=1.20$, $\varphi_0=0$, $\psi=0.37$ (upper plot): solid and dashed lines (blue and orange) show its real and imaginary parts. Below, the response curves at $\omega_0\tau=9.03$ for $\eta=0,0.02,0.04$ span $h=0.5$--$1.66$; only responses up to 5.2 are displayed. The common $n=2$ plateau persists while its boundaries move. Finite-$q$ crossings approximate the irrational bulk boundaries (Supplemental Material, Secs.~S3\,A and S6\,A).}
\label{fig:three}
\end{figure}

The asymptotic logarithmic amplification per layer is characterized by the
Lyapunov exponent~\cite{Avila2015,AvilaJitomirskayaSadel2014}. We denote
this bulk growth rate by $g_\infty$:
\begin{equation}
g_\infty(h)=\lim_{N\to\infty}\frac1N
\int_0^1\ln\|M_N(\varphi,h)\|_2\,d\varphi.
\label{eq:ginf}
\end{equation}
Here $\|\cdot\|_2$ is the spectral norm; the output of a finite sequence for a given input is treated in the Supplemental Material, Sec.~S5.
For irrational $\alpha$, Avila's acceleration theorem gives
\begin{equation}
\partial_h^+g_\infty=n\in\mathbb{Z},
\label{eq:headline}
\end{equation}
where the superscript $+$ denotes the right derivative~\cite{Avila2015,AvilaJitomirskayaSadel2014}.
The integer $n$ characterizes the response to complex-phase continuation. Inside each amplifying phase, it is a winding number: as the modulation phase goes through one period, the complex factor by which one layer multiplies the growing solution winds $n$ times around the origin (Supplemental Material, Sec.~S2\,A). This topological invariant is distinct from the gap Chern numbers of the Hermitian PTQC~\cite{Ni2025}.
The permittivity varies smoothly with $h$, and the growth rate remains
continuous. Its derivative, however, forms plateaus at integer values.
Within each such interval, $g_\infty(h)=b+nh$, with $b$ constant.
Where the integer changes, the growth curve therefore has a corner.

Figure~\ref{fig:two}(a) shows the bulk growth rate at fixed
$\omega_0\tau$, and Fig.~\ref{fig:two}(b) resolves its integer slope (numerical methods and convergence checks in the Supplemental Material, Sec.~S6\,A). Colors identify the successive integer phases.

Figure~\ref{fig:two}(c) maps these integer phases in the
$(h,\omega_0\tau)$ plane. Within the analytic strip and away from
phase boundaries, this phase diagram displays the local rigidity of the
response, in that continuous parameter changes within a connected integer phase
leave $n$ unchanged even as the amplification rate varies.
Crossing a phase boundary changes the integer response, whose transition
mechanism is examined below. Different input frequencies can therefore
belong to different integer phases under the same modulation, providing
the basis for the spectral selection discussed later.

To locate the transitions between integer phases, we apply the rational-approximation framework underlying Avila's proof of acceleration quantization~\cite{Avila2015}. We choose the continued-fraction convergents $p/q$ of the irrational modulation frequency; for the golden mean, their denominators are Fibonacci numbers. Here $q$ labels an auxiliary periodic approximant, not the length of the propagated sequence. The $q$-layer trace of this approximant, in which $\alpha$ is replaced by $p/q$, is periodic in $\varphi$ with period $1/q$. We define $C_n^{(q)}(h)$ as the complex amplitude of the Fourier harmonic $e^{-2\pi\ii nq\varphi}$ in this trace,
\begin{equation}
C_n^{(q)}(h)=q\int_0^{1/q}\operatorname{tr}M_q(\varphi,h)\,e^{2\pi\ii nq\varphi}\,d\varphi.
\label{eq:Fourier}
\end{equation}
The trace and these coefficients therefore satisfy
\begin{equation}
\begin{aligned}
\operatorname{tr}M_q(\varphi,h)&=\sum_{n\in\mathbb Z}C_n^{(q)}(h)\,e^{-2\pi\ii nq\varphi},\\
q^{-1}\ln|C_n^{(q)}(h)|&=b_n^{(q)}+nh,
\end{aligned}
\label{eq:lines}
\end{equation}
where $b_n^{(q)}=q^{-1}\ln|C_n^{(q)}(0)|$. The second relation follows directly from analytic continuation, within the analytic strip. Expanding each layer matrix in harmonics $e^{-2\pi\ii k\varphi}$ of the modulation phase, complexification weights the harmonic of order $k$ by $e^{kh}$, as an imaginary gauge field weights a hop of $k$ sites in the Hatano--Nelson model~\cite{HatanoNelson1996}. The coefficient $C_n^{(q)}$ collects the terms of the $q$-layer product whose orders add up to $nq$, that is, $n$ net harmonic steps per layer, and is therefore weighted by $e^{nqh}$. Each Fourier order thus contributes a straight line of integer slope. Because each term scales exponentially with $q$, the largest one dominates the trace, so the bulk growth rate follows the highest line, or vanishes if all lines are negative, and inherits its integer slope~\cite{Avila2015}. A transition between integer phases is an exchange of dominant orders, located where adjacent lines cross (Supplemental Material, Sec.~S3\,A),
\begin{equation}
h_{c,n}^{(q)}=b_n^{(q)}-b_{n+1}^{(q)}.
\label{eq:crossing}
\end{equation}

Figure~\ref{fig:three}(a) shows, as a representative example, orders $n=3,4,5$ dominating in succession, together with an independent numerical estimate of the bulk response. The crossings approximate the corners separating the integer phases. Figure~\ref{fig:three}(b) makes the change of dominant order visible through normalized coefficient magnitudes. Figure~\ref{fig:three}(c) follows the predicted boundaries to higher orders at three input frequencies. The resolved high-order phase boundaries crowd toward $h_\star$, where the analytically continued permittivity first vanishes. Despite the increasing integer response, the bulk growth rate remains bounded and approaches a finite limit as $h\to h_\star^-$ (Supplemental Material, Sec.~S3\,B).

The staircase of integer phases has no counterpart in the spatial non-Hermitian AAH model, whose transfer matrix contains only the harmonics of orders $0$ and $\pm1$, so that the integer there is at most one and its single transition has a closed form~\cite{Longhi2019a,Liu2021,Avila2015}. Here each layer lasts a finite time, over which the wave accumulates the phase $\omega_0\tau z_m$, so the layer matrix contains harmonics of all orders and the integer can exceed one. Near $h_\star$, the permittivity of some layers approaches zero, and the wave oscillates ever faster within them. The integer $n$ then approximately counts the half periods that the wave completes within the layer of smallest permittivity (Supplemental Material, Sec.~S3\,B). Equation~\eqref{eq:crossing} nevertheless predicts each exchange of dominant orders explicitly, from two Fourier coefficients at a single value of $h$, without fitting or computing the growth rate. As the input frequency changes, the predicted boundaries shift [Fig.~\ref{fig:three}(d)] and agree closely with those obtained by direct transfer-matrix propagation of the irrational sequence (Supplemental Material, Sec.~S3\,A). Carrier-envelope pseudospectral time-domain (PSTD) simulations~\cite{Liu1997PSTD,Chen2008PSTD}, which integrate Maxwell's equations in space and time, agree with them as well (Supplemental Material, Sec.~S6\,B).

To test the stability of the integer phases, we distort the permittivity modulation by a second harmonic, analytically continued in the same way,
\begin{equation}
\begin{aligned}\varepsilon_m^{(\eta)}&=\varepsilon_R+\delta\varepsilon\sin Z_m+\eta\,\delta\varepsilon\sin(2Z_m+\psi),\\ Z_m&=X_m+\ii h,\end{aligned}
\label{eq:distortion}
\end{equation}
where $\eta$ measures the distortion amplitude and $\psi$ is its fixed phase. This distortion alters both the real and imaginary parts of the permittivity throughout the temporal sequence [upper plot of Fig.~\ref{fig:three}(e)]. The bulk response nevertheless retains its integer plateaus, as the common $n=2$ plateau at $h=1.20$ shows [Fig.~\ref{fig:three}(e)]. Their boundaries move with the distortion strength, while the integers they carry stay the same.

This persistence follows from uniform hyperbolicity, the exponential separation of growing and decaying solutions at every modulation phase, which holds inside every amplifying integer phase (Supplemental Material, Sec.~S2\,A). There the acceleration is a winding number in the modulation phase~\cite{Avila2015,AvilaJitomirskayaSadel2014}, a topological integer. A continuous waveform distortion that preserves the analytic continuation and uniform hyperbolicity therefore cannot change it (Supplemental Material, Sec.~S4\,A).

\begin{figure}[t]
\centering
\includegraphics[width=\columnwidth]{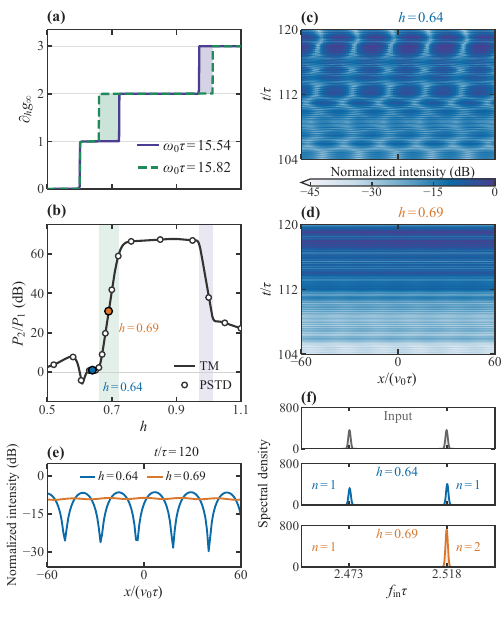}%
\caption{\textbf{Integer response and spectral amplification.}
(a) Integer slopes $\partial_h g_\infty$ at $\omega_{0,1}\tau=15.54$ and $\omega_{0,2}\tau=15.82$. Green and purple fills mark where the second and first band, respectively, has the larger integer.
(b) Output band-power ratio $10\log_{10}(P_2/P_1)$ for a fixed, equal-power two-band input, $N=120$, and $\varphi_0=0$: transfer-matrix prediction (TM, line) and PSTD simulation (circles). Matching shaded intervals identify increasing and decreasing relative amplification. Colored circles select $h=0.64$ (blue) and 0.69 (orange).
(c,d) PSTD displacement-field intensities at these two settings. Each map displays $10\log_{10}[|D(x,t)|^2/\max_{x,t}|D|^2]$, normalized to its own maximum over all saved positions and times; the shared color scale compares relative field structure, not absolute gain between runs; $v_0=c/\sqrt{\varepsilon_R}$.
(e) Final spatial profiles with the same respective normalizations.
(f) Input and output power spectra, each normalized to unit integral. The frequency axis labels spatial modes by their reference frequencies, $f_{\rm in}\tau=\omega_0\tau/(2\pi)$; the peak labels give the bulk integers at the band centers.}
\label{fig:four}
\end{figure}

When two input frequencies lie in different integer phases of the same modulation, their integer difference determines how sensitively their relative amplification responds to the complexification parameter $h$. We launch a coherent input with equal power in narrow bands centered at $\omega_{0,1}\tau=15.54$ and $\omega_{0,2}\tau=15.82$. Let $P_j$ denote the output power after $N$ layers, integrated over band $j$ and summed over both propagation directions in the real reference medium. For bands lying within integer phases and an input coupled to the growing solutions, the leading power amplification $\exp(2Ng_j)$, with $g_j$ the bulk growth rate at the center of band $j$, gives
\begin{equation}
\frac{\partial}{\partial h}\left[10\log_{10}\frac{P_2}{P_1}\right]\simeq\frac{20N}{\ln 10}\,[n_2(h)-n_1(h)].
\label{eq:ratio}
\end{equation}
Here $n_j=\partial_h g_j$. Finite length and bandwidth introduce corrections (Supplemental Material, Secs.~S5\,A and S5\,B). Equation~\eqref{eq:ratio} applies to a single sequence because $\alpha$ is irrational. The phases $X_m$ of one sequence then never repeat and fill the modulation cycle uniformly for any initial phase, so one long sequence performs the phase average of Eq.~\eqref{eq:ginf} by itself, and inside an amplifying phase its growth approaches the bulk rate. A periodic sequence visits only a finite set of phases, and a random sequence is not generated by an advancing phase at all. Neither gives integer plateaus for a single sequence (Supplemental Material, Sec.~S4\,B). Within an interval where the two integers differ by one, the relative output power varies exponentially with $h$, with a logarithmic sensitivity proportional to the propagation length.

The offset between the two integer staircases in Fig.~\ref{fig:four}(a) thus determines the rise, plateau, and fall of the power ratio in Fig.~\ref{fig:four}(b). In the green interval, the second band has entered $n=2$ while the first remains at $n=1$, so increasing $h$ amplifies the second band more strongly and the output power ratio rises steeply. When the first band also enters $n=2$, both growth rates acquire the same slope. Further increases in $h$ then preserve the large output spectral contrast to leading order, producing the broad plateau. The first band subsequently reaches $n=3$ before the second. Their integer difference changes sign in the purple interval, and the power ratio falls. Because the rise, plateau, and fall depend on where the phase boundaries lie, the waveform distortions discussed above can serve for spectral control. By shifting the boundaries while keeping the integers [Fig.~\ref{fig:three}(e); Supplemental Material, Sec.~S5\,B], they tune the intervals over which the power ratio rises, stays flat, or falls.

For the same equal-power input and initial modulation phase, changing $h$ from 0.64 to 0.69 changes the phase pair from $(1,1)$ to $(1,2)$, with the second band entering $n=2$ at $h\approx0.660$, and increases the output power ratio by approximately three orders of magnitude after 120 layers. PSTD integration of the first-order Maxwell equations on a spatial grid independently reproduces the transfer-matrix curve [Fig.~\ref{fig:four}(b)]. The resulting dominance of the second band is visible both in the output spectrum [Fig.~\ref{fig:four}(f)] and in the suppression of the pronounced spatial beating between the two bands, the interference pattern of the propagated field [Fig.~\ref{fig:four}(c--e)]. The bulk quantization thus leaves a measurable imprint on the propagated wave. Within an interval of fixed integers, the slope of the power ratio, normalized as in Eq.~\eqref{eq:ratio}, approaches the integer difference as the sequence lengthens and already separates differences of one and zero after 120 layers (Supplemental Material, Sec.~S5\,B). The integer difference can therefore be extracted from a scan of the power ratio.

\makeatletter
\@booleantrue\raggedcolumn@sw
\makeatother
\raggedbottom
\emph{Conclusion.---}We have shown that amplification in a non-Hermitian photonic time quasicrystal responds to the complexification parameter with integer slopes. The permittivity and the growth rate both vary continuously with this parameter, yet the slope of the growth rate can take only integer values, given by Avila's acceleration, and a single-harmonic modulation already drives it through a staircase of successive integers. These integers are topological winding numbers. Analytic distortions of the modulation shift the boundaries between them but cannot change them, and Fourier coefficients of rational approximants predict where the boundaries lie. For a wave whose two frequency components lie in different integer phases, the difference of their integers sets how fast the output power ratio between them varies with $h$, which yields spectral selection.

Here non-Hermiticity acts as an imaginary gauge field on the harmonics of the modulation, which links the quantized response to non-Hermitian physics in synthetic frequency dimensions~\cite{Yuan2018}. Topology thereby makes the response of amplification to this imaginary gauge field quantized, through an integer of a different kind from the gap Chern numbers. Mathematically, the mechanism requires only a wave equation whose coefficient is modulated in temporal layers. The same quantized response should therefore arise in principle in acoustic and elastic waves and in electrical circuits, and a transmission line built from lumped circuit elements implements the model exactly (Supplemental Material, Sec.~S7). The quantized response of amplification could be exploited for reconfigurable gain control, frequency-selective amplification, and ratiometric sensing in optical and microwave devices.

\par\vspace*{6pt}
\begin{acknowledgments}
This work was supported by the National Natural Science Foundation of China
(No. 62201541).
\end{acknowledgments}

The data that support the findings of this article are available from the authors upon reasonable request.

\clearpage
\section*{Supplemental Material}
\addcontentsline{toc}{section}{Supplemental Material}
\setcounter{section}{0}
\setcounter{figure}{0}
\setcounter{table}{0}
\setcounter{equation}{0}
\renewcommand{\thesection}{S\arabic{section}}
\renewcommand{\thefigure}{S\arabic{figure}}
\renewcommand{\thetable}{S\arabic{table}}
\renewcommand{\theequation}{S\arabic{equation}}
\newcommand{\dd}{\mathrm{d}}
\newcommand{\wind}{\operatorname{wind}}
\linespread{0.96}
\setlength{\abovedisplayskip}{6pt plus 2pt minus 2pt}
\setlength{\belowdisplayskip}{6pt plus 2pt minus 2pt}
\setlength{\textfloatsep}{7pt plus 2pt minus 2pt}
\setlength{\intextsep}{7pt plus 2pt minus 1pt}
\setlength{\abovecaptionskip}{5pt}
\setlength{\floatsep}{8pt plus 2pt minus 1pt}
\setlength{\dbltextfloatsep}{13pt plus 2pt minus 2pt}
\setlength{\dblfloatsep}{10pt plus 2pt minus 2pt}
\renewcommand{\topfraction}{0.95}
\renewcommand{\bottomfraction}{0.95}
\renewcommand{\textfraction}{0.05}
\renewcommand{\floatpagefraction}{0.75}
\setcounter{topnumber}{3}
\setcounter{totalnumber}{4}
\raggedbottom

\begin{center}
\textbf{Supplemental Material for ``Quantized response of amplification in non-Hermitian photonic time quasicrystals''}
\end{center}
This Supplemental Material gives the derivations and numerical support for
the conclusions of the main text. Section~S1 derives the layer matrix from
Maxwell's equations and introduces the complexified modulation; Sec.~S2
proves that the response of the growth rate to $h$ is an integer; Sec.~S3
predicts the phase boundaries from the Fourier coefficients of rational
approximants and gives the law of their accumulation at the edge of the
analytic strip; Sec.~S4 explains under which perturbations the integer
response is preserved and why quasiperiodicity is needed; Sec.~S5 discusses
the output of finite sequences and two-frequency inputs; Sec.~S6 gives the
numerical methods and the scope of the model; Sec.~S7 gives a circuit that
realizes the model of this work.

\section{Temporal layers as an analytic cocycle}

\subsection{Maxwell's equations and the layer matrix}

Consider a spatially uniform medium whose permeability is the vacuum value
$\mu_0$ and whose relative permittivity $\varepsilon(t)$ varies only in
time. A plane wave propagates along $x$ with its electric field along $y$,
so the nonzero field components are $E_y$, $D_y$, and $B_z$. With the
instantaneous constitutive relation $D_y=\varepsilon_0\varepsilon(t)E_y$,
Maxwell's equations are
\begin{equation}
\partial_x E_y=-\partial_t B_z,\qquad
-\mu_0^{-1}\partial_xB_z=\partial_tD_y.
\label{eq:S1}
\end{equation}
The coefficients of the equations do not depend on $x$, so the medium is
invariant under spatial translation and the wavenumber $k$ is conserved.
The equations are also linear, so different wavenumbers do not mix, and
each Fourier component $\ee^{\ii kx}$ evolves on its own. For the
amplitudes $D_k$, $E_k$, and $B_k$ of one component, eliminating $B_k$ and
inserting the constitutive relation gives a closed equation for the
displacement amplitude,
\begin{equation}
\ddot D_k=-\frac{k^2}{\mu_0}E_k=-\frac{c^2k^2}{\varepsilon(t)}D_k,
\qquad c^2=(\varepsilon_0\mu_0)^{-1}.
\label{eq:S2}
\end{equation}

We take a real, positive reference permittivity $\varepsilon_R$. The
modulation introduced below oscillates about $\varepsilon_R$; before the
modulation starts and after it ends, the permittivity of the medium is
$\varepsilon_R$. We label each Fourier component by its frequency in this
reference medium, $\omega_0=c|k|/\sqrt{\varepsilon_R}$. We introduce the
dimensionless time $s=\omega_0t$ and write the ratio of the reference
permittivity to the instantaneous permittivity as
$z(s)^2=\varepsilon_R/\varepsilon(s)$. Below, a prime always denotes a
derivative with respect to $s$, $D'\equiv\dd D/\dd s$, where $D\equiv D_k$;
$D'$ has the same dimension as $D$. Equation~\eqref{eq:S2} then reads
$D''=-z^2D$. The values of $D$ and $D'$ at any one time determine the whole
solution. We combine these two quantities into the state
$v=(D,\ii D')^{\mathsf T}$ of the main text. The factor $\ii$ is added so
that the two waves of the reference medium ($z=1$) correspond to simple
states: a forward wave $D\propto\ee^{-\ii s}$ corresponds to
$v\propto(1,1)^{\mathsf T}$, and a backward wave $D\propto\ee^{\ii s}$
corresponds to $v\propto(1,-1)^{\mathsf T}$. In terms of the state,
Eq.~\eqref{eq:S2} becomes
\begin{equation}
\frac{\dd v}{\dd s}=\begin{pmatrix}0&-\ii\\-\ii z(s)^2&0\end{pmatrix}v.
\label{eq:S3}
\end{equation}

At a time when $\varepsilon$ switches, the fields remain finite.
Integrating Eq.~\eqref{eq:S1} over an infinitesimal interval that contains
this time shows that $D_k$ and $B_k$ are both continuous. The second
equation of~\eqref{eq:S1} gives $D'=-\ii kB_k/(\mu_0\omega_0)$; the
permeability is not modulated and $\mu_0$ is a constant, so $D'$ is also
continuous. Therefore the state $v$ takes the same value on both sides of
every interface. In contrast, the electric field
$E_k=D_k/(\varepsilon_0\varepsilon)$ jumps together with the permittivity;
this is why we use $D$ rather than $E$. As a result, propagation through a
sequence of layers is the product of the matrices that carry $v$ through
each layer, and the interfaces need no separate matrices.

Let layer $m$ ($m=0,\ldots,N-1$) occupy the time interval
$m\tau\le t<(m+1)\tau$ with a constant permittivity $\varepsilon_m$, and
write $z_m^2=\varepsilon_R/\varepsilon_m$. The dimensionless duration of
each layer is $\omega_0\tau$. Inside a layer $z=z_m$ is constant, so the
solutions of Eq.~\eqref{eq:S3} are the two waves
$D\propto\ee^{\mp\ii z_ms}$, with local frequencies $\pm\omega_0z_m$.
Integrating Eq.~\eqref{eq:S3} over one layer gives the matrix that carries
$v$ through this layer, which is the layer matrix of the main text.
Propagation through $N$ layers is the ordered product of these matrices:
\begin{equation}
\begin{gathered}
S_m=\begin{pmatrix}\cos(\omega_0\tau z_m)&-\ii\sin(\omega_0\tau z_m)/z_m\\
-\ii z_m\sin(\omega_0\tau z_m)&\cos(\omega_0\tau z_m)\end{pmatrix},\\ M_N=S_{N-1}\cdots S_0.
\end{gathered}
\label{eq:S4}
\end{equation}
A direct calculation from Eq.~\eqref{eq:S4} gives
$\det S_m=\cos^2(\omega_0\tau z_m)+\sin^2(\omega_0\tau z_m)=1$, so each
$S_m$ and the product $M_N$ belong to $\mathrm{SL}(2,\mathbb C)$, and the
two eigenvalues of $M_N$ have product 1. Growth is therefore always
accompanied by decay: when one eigenvalue has modulus larger than 1, the
other must have modulus smaller than 1. In a photonic time crystal with
real permittivity, $M_N$ is the $N$th power of the matrix of one
modulation period; inside a momentum gap, the eigenvalue $\mu$ of this
matrix satisfies $|\mu|>1$, and the two eigenvalues of $M_N$, $\mu^N$ and
$(1/\mu)^N$, grow and decay exponentially, respectively~\cite{Lyubarov2022}.

We can also show that $S_m$ is a single-valued holomorphic function of
$\varepsilon_m$ as long as $\varepsilon_m\neq0$. The layer matrix contains
the square root $z_m=(\varepsilon_R/\varepsilon_m)^{1/2}$, which has two
branches when $\varepsilon_m$ is complex, but the choice of branch does not
matter: replacing $z_m$ by $-z_m$ only exchanges the two waves
$\ee^{\mp\ii z_ms}$ in the layer, and the layer matrix propagates both
waves. In terms of formulas, the three matrix elements of $S_m$,
$\cos(\omega_0\tau z_m)$, $\sin(\omega_0\tau z_m)/z_m$, and
$z_m\sin(\omega_0\tau z_m)$, are all even functions of $z_m$ and therefore
depend only on $z_m^2=\varepsilon_R/\varepsilon_m$. They are also power
series in $z_m^2$ that converge everywhere, for example
$\cos(\omega_0\tau z_m)=\sum_{\ell\ge0}(-\omega_0^2\tau^2\varepsilon_R/\varepsilon_m)^\ell/(2\ell)!$,
and are thus holomorphic functions of $\varepsilon_m$ wherever
$\varepsilon_m\neq0$.

\subsection{Complexified modulation and the cocycle}

The permittivity of layer $m$ is set by the modulation phase
$X_m=2\pi(\varphi_0+m\alpha)$. Here $\varphi_0\in[0,1)$ is the initial
phase; the irrational number $\alpha$ is called the rotation number, and it
sets how far the phase advances from one layer to the next. With the
sinusoidal modulation $\varepsilon_m=\varepsilon_R+\delta\varepsilon\sin X_m$
and an amplitude $0<\delta\varepsilon<\varepsilon_R$, such a medium is a
Hermitian photonic time quasicrystal~\cite{Ni2025}. We continue the
modulation phase analytically into the complex plane,
$X_m\to X_m+\ii h$. This defines the complexification parameter
$h\in\mathbb R$, and the permittivity becomes
\begin{equation}
\varepsilon_m=\varepsilon_R+\delta\varepsilon\sin(X_m+\ii h).
\label{eq:S5}
\end{equation}
For $h\neq0$ the permittivity is complex. Time-periodic modulation of a
complex permittivity has been studied before in photonic time
crystals~\cite{Wang2018}. Unless stated otherwise, the parameters are those
of the main text: $\varepsilon_R=3.4$, $\delta\varepsilon=1$,
$\alpha=(\sqrt5-1)/2$ (the golden mean), and numerical examples use the
frequency $\omega_0\tau=9.03$; the figure captions state it again. The real
and imaginary parts are
\begin{equation}
\begin{gathered}
\operatorname{Re}\varepsilon_m=\varepsilon_R+\delta\varepsilon\cosh h\,\sin X_m,\\
\operatorname{Im}\varepsilon_m=\delta\varepsilon\sinh h\,\cos X_m.
\end{gathered}
\label{eq:S6}
\end{equation}
They oscillate from layer to layer with a quarter-period phase difference,
and their amplitudes $\delta\varepsilon\cosh h$ and $\delta\varepsilon\sinh h$
are both set by the same parameter $h$.

As $|h|$ increases from zero, the layer matrix remains analytic until the
permittivity vanishes at some phase (Sec.~S1\,A). From
$\sin(X+\ii h)=\sin X\cosh h+\ii\cos X\sinh h$, the first zero appears at
$X=3\pi/2$, when $|h|$ equals
\begin{equation}
h_\star=\operatorname{arccosh}(\varepsilon_R/\delta\varepsilon).
\label{eq:S7}
\end{equation}
For $\varepsilon_R=3.4$ and $\delta\varepsilon=1$, $h_\star=1.8946$. The
layer matrix is therefore analytic for $|h|<h_\star$, and we call this
range the analytic strip. Below we consider only $h$ inside the strip;
inside the strip,
$\operatorname{Re}\varepsilon_m\ge\varepsilon_R-\delta\varepsilon\cosh h>0$.

Write the modulation phase of one layer as $X=2\pi\varphi$, so that
$X+\ii h=2\pi(\varphi+\ii h/2\pi)$. Thus the layer matrix does not depend
on $\varphi$ and $h$ separately, but only on their combination
$\vartheta=\varphi+\ii h/2\pi$, which we call the complex phase. Let
$S(\vartheta)$ denote the layer matrix at the complex phase $\vartheta$,
and write $S_h(\varphi)=S(\varphi+\ii h/2\pi)$. In each layer, the phase
advances by $\alpha$ and the state vector is multiplied by the layer matrix
of this layer: $(\varphi,v)\mapsto(\varphi+\alpha,\,S_h(\varphi)v)$. After
$N$ layers,
\begin{equation}
M_N(\varphi,h)=S_h\bigl(\varphi+(N-1)\alpha\bigr)\cdots S_h(\varphi),
\label{eq:S8}
\end{equation}
and $\varphi=\varphi_0$ gives back the layer sequence of
Eq.~\eqref{eq:S5}. The two rules, that the phase advances by $\alpha$ in
each layer and that the state vector is multiplied by $S_h(\varphi)$ in
each layer, together are called a cocycle. In the classification of
quasiperiodic operator theory~\cite{Avila2015,AvilaJitomirskayaSadel2014},
it is an analytic one-frequency $\mathrm{SL}(2,\mathbb C)$ cocycle: the
layer matrix is an analytic function of the phase and is set by a single
phase $\varphi$, which advances by the same angle $\alpha$ in every layer;
if the modulation were set jointly by several mutually incommensurate
phases, the cocycle would be called multifrequency. The theorems of Avila
used in this work apply to exactly this class of one-frequency cocycles.

Fixing the initial phase $\varphi_0$ and the number of layers $N$ gives a
finite sequence, which we call a single sequence below. Its growth rate per
layer is $N^{-1}\ln\|M_N(\varphi_0,h)\|$, where $\|\cdot\|$ is the
spectral norm. This quantity changes with $N$ and may also depend on the
initial phase. To obtain a quantity that belongs to the medium alone and
does not depend on the initial phase, we first average over the initial
phase and then let $N\to\infty$:
\begin{equation}
g_\infty(h)=\lim_{N\to\infty}\frac1N\int_0^1\ln\|M_N(\varphi,h)\|\,\dd\varphi.
\label{eq:S9}
\end{equation}
For each $h$, $g_\infty(h)$ is a definite number; it is the bulk growth
rate of the main text and is called the Lyapunov exponent in cocycle
theory. The limit exists: the norm of a product does not exceed the
product of the norms of its parts, and in the average over the initial
phase a shift of the phase by $N\alpha$ does not change the average, so
$a_N=\int_0^1\ln\|M_N\|\,\dd\varphi$ satisfies $a_{N+K}\le a_N+a_K$
(subadditivity), and therefore $a_N/N$ converges. The average over the
initial phase is only a convenience of the definition; a single sequence
does not need it. The phases that a single sequence passes through in turn
are $\varphi_0,\varphi_0+\alpha,\varphi_0+2\alpha,\ldots$; for irrational
$\alpha$ these phases are uniformly distributed on $[0,1)$. If
$\ln\|M_N\|$ were a sum of contributions of the individual layers, the
average along the sequence would equal the average over the phase, and the
growth rate of a single sequence would naturally tend to $g_\infty(h)$. In
fact $\ln\|M_N\|$ is only subadditive and is not a sum over layers;
Kingman's subadditive ergodic theorem~\cite{Kingman1968} guarantees that
the conclusion still holds in this case: for almost every initial phase
$\varphi_0$, $N^{-1}\ln\|M_N(\varphi_0,h)\|$ tends to $g_\infty(h)$ as
$N\to\infty$.

\section{Integer response}

\subsection{Acceleration and the winding of the multiplier}

The real part of the complex phase $\vartheta=\varphi+\ii h/2\pi$ is the
phase $\varphi$, and the imaginary part is $h/2\pi$. At fixed $N$, $M_N$ is
a holomorphic function of $\vartheta$, so $\ln\|M_N\|$ is a subharmonic
function of $\vartheta$: its value at any point does not exceed its average
over a small circle around that point. Averaging it over the phase
$\varphi$ over one period gives the phase-averaged log norm per layer,
$N^{-1}\int_0^1\ln\|M_N\|\,\dd\varphi$. Since $\varphi$ has been averaged
out, this quantity depends only on the imaginary part, that is, only on
$h$, and it is still subharmonic. A subharmonic function that depends on
only one real variable is a convex function, so it is a convex function of
$h$; its limit $g_\infty$ as $N\to\infty$ [Eq.~\eqref{eq:S9}] is also a
convex function of $h$~\cite{Avila2015}. A convex function has a right
derivative at every point, and where the slope does not jump, the right
derivative is the ordinary derivative. Avila takes the imaginary part
$y=h/2\pi$ as the variable, writes the Lyapunov exponent as $L(y)$, and
defines $(2\pi)^{-1}\partial_y^+L$ as the acceleration, where
$\partial_y^+$ is the right derivative. In his global theory he proved that
the acceleration is an integer when $\alpha$ is
irrational~\cite{Avila2015,AvilaJitomirskayaSadel2014}. Since
$L(y)=g_\infty(2\pi y)$, the factor $2\pi$ cancels when we change to our
variable $h$, and the acceleration is the right derivative of $g_\infty$
with respect to $h$:
\begin{equation}
n(h)=\partial_h^+g_\infty(h)\in\mathbb Z.
\label{eq:S10}
\end{equation}

The bulk growth rate $g_\infty$ is an even function of $h$. The reason is as follows. The
coefficients of the power series of $\sin$ are all real and $X$ is real, so
the complex conjugate of $\sin(X+\ii h)$ is $\sin(X-\ii h)$; that is, the
complex conjugate of the permittivity at parameter $h$ equals the
permittivity at parameter $-h$. The diagonal elements of the layer matrix
are series in $1/\varepsilon_m$ with real coefficients, and the two
off-diagonal elements are such series multiplied by $-\ii$; therefore
taking the complex conjugate of the layer matrix is the same as replacing
$h$ by $-h$ and changing the sign of the off-diagonal elements,
$\overline{S_h(\varphi)}=\sigma_zS_{-h}(\varphi)\sigma_z$, where
$\sigma_z=\operatorname{diag}(1,-1)$. Neither complex conjugation nor
multiplication by $\sigma_z$ on both sides changes the norm, so
$g_\infty(h)=g_\infty(-h)$. On the other hand, the norm of the layer matrix
is bounded inside the analytic strip, so $g_\infty$ is finite; a finite
convex function is automatically continuous on an open interval. A
continuous, even, convex function takes its minimum at $h=0$, and for
$h>0$ its slope is nonnegative and does not decrease. On any closed
interval that does not reach the edge of the strip, the slope is bounded
and can only take integer values, so it takes only finitely many values.
The growth rate therefore consists of straight segments: on each interval
where the acceleration $n$ stays constant, $g_\infty(h)=b+nh$, where the
slope $n$ of this linear function is the integer acceleration of
Eq.~\eqref{eq:S10}, and $b$ is the intercept of this function at $h=0$.
These intervals are the integer phases of the main text. Since $g_\infty$
is continuous, two adjacent segments meet where their intervals join, and
the intercept $b$ changes from one integer phase to the next.

In an integer phase with $g_\infty>0$, the propagation has a simple
structure called uniform hyperbolicity. Describing it requires the notion
of a ``direction''. Any state can be written as a superposition of the
forward and backward waves of the reference medium,
$v=A_+(1,1)^{\mathsf T}+A_-(1,-1)^{\mathsf T}$, with complex coefficients
$A_+$ and $A_-$. Multiplying $v$ by a nonzero complex number multiplies
$A_+$ and $A_-$ by the same number, and their ratio $A_+:A_-$ does not
change. States with the same ratio have the same mixture of forward and
backward waves and differ only in overall amplitude and phase; we say that
they belong to the same direction. For a periodic modulation, the two
eigenvectors of the matrix of one period give two special directions: a
wave composed with one of these mixtures is amplified by the same factor in
every period, and this is the growing direction; a wave composed with the
other mixture is reduced by the same factor in every period, and this is
the decaying direction. A quasiperiodic modulation also has two such
directions, except that they vary with the phase. Uniform hyperbolicity
means that at every phase $\varphi$ the state space splits into a growing
direction $E^u(\varphi)$ and a decaying direction $E^s(\varphi)$. Through
each layer, the growing direction at phase $\varphi$ is mapped to the
growing direction at phase $\varphi+\alpha$, and the decaying direction is
mapped to the decaying direction, so the two kinds of directions do not
turn into each other. The two directions are always separated by a nonzero
angle. Vectors in the decaying direction shrink at a uniform exponential
rate, and vectors in the growing direction grow at a uniform exponential
rate. We write the latter with the inverse product instead: the inverse
product shrinks vectors in the growing direction at the final phase.
Together,
\begin{equation}
\begin{gathered}
\|M_N(\varphi)v_s\|\le C_0\ee^{-\gamma N}\|v_s\|,\\
\|M_N(\varphi)^{-1}v_u\|\le C_0\ee^{-\gamma N}\|v_u\|,
\end{gathered}
\label{eq:S11}
\end{equation}
which holds for all $\varphi$, all $N\ge0$, all $v_s\in E^s(\varphi)$, and
all $v_u\in E^u(\varphi+N\alpha)$, where $C_0,\gamma>0$ are constants.
Avila proved that, when the Lyapunov exponent is positive, if $g_\infty$
is a straight segment near some $h$, then the cocycle is uniformly
hyperbolic there, and the converse also holds~\cite{Avila2015}. The growth
rate is a straight segment in every integer phase, so the cocycle is
uniformly hyperbolic in every integer phase with $g_\infty>0$. Because the
cocycle is holomorphic, its growing and decaying directions then depend
analytically on the complex phase~\cite{AvilaJitomirskayaSadel2014}.

How much a vector in the growing direction is amplified, and how much it
turns, in each layer can be described by one complex number. For this
purpose we choose, at every complex phase $\vartheta$, a nonzero vector
$u(\vartheta)$ that points in the growing direction. It lies in the same
two-dimensional state space as the state $v$, but it is not a particular
state in the propagation; it is a reference vector chosen for each phase,
and we require it to be a periodic holomorphic function of $\vartheta$.
The growing direction itself depends analytically and periodically on
$\vartheta$, so such a vector can always be
found~\cite{AvilaJitomirskayaSadel2014}. One layer maps the growing
direction to the growing direction, so it maps $u(\vartheta)$ to a multiple
of $u(\vartheta+\alpha)$. We call this complex factor the multiplier
$\lambda(\vartheta)$:
\begin{equation}
S(\vartheta)u(\vartheta)=\lambda(\vartheta)u(\vartheta+\alpha),\qquad
\lambda(\vartheta)\ne0.
\label{eq:S12}
\end{equation}
Applying $N$ layers in succession multiplies the multipliers layer by
layer, and taking the logarithm gives
\begin{equation}
\ln\frac{\|M_Nu(\vartheta)\|}{\|u(\vartheta)\|}
=\sum_{m=0}^{N-1}\ln|\lambda(\vartheta+m\alpha)|
+\ln\frac{\|u(\vartheta+N\alpha)\|}{\|u(\vartheta)\|},
\label{eq:S13}
\end{equation}
where the last term is bounded uniformly in $\varphi$. The two directions
are always separated by a nonzero angle, so, up to a bounded factor, the
largest amplification of a vector by a long product is achieved by a
vector in the growing direction. Therefore $\|M_Nu\|/\|u\|$ lies between
two constant multiples of $\|M_N\|$, and the two constants depend on
neither $\varphi$ nor $N$. We divide by $N$, average over $\varphi$, and
let $N\to\infty$: the bounded term divided by $N$ tends to zero; each term
in the sum of multipliers gives the same value after the phase average,
because a shift of the phase does not change the average. The growth rate
is therefore the average of the log modulus of the multiplier,
\begin{equation}
g_\infty(h)=\int_0^1\ln\bigl|\lambda(\varphi+\ii h/2\pi)\bigr|\,\dd\varphi.
\label{eq:S14}
\end{equation}

From Eq.~\eqref{eq:S14} the acceleration can be written as the winding
number of the multiplier. The multiplier $\lambda$ is a holomorphic function of
$\vartheta$ and $\vartheta=\varphi+\ii h/2\pi$, so the derivative of
$\lambda$ with respect to $h$ equals its derivative with respect to
$\varphi$ multiplied by $\ii/2\pi$ (Cauchy--Riemann relations).
Differentiating Eq.~\eqref{eq:S14}, the integrand becomes
$\partial_h\ln|\lambda|=-(2\pi)^{-1}\partial_\varphi\arg\lambda$, the rate
of change of the argument of the multiplier along $\varphi$. Integrating
over one period of $\varphi$ gives the total change of the argument over
one period divided by $-2\pi$:
\begin{equation}
\partial_hg_\infty=-\frac{1}{2\pi}\int_0^1\partial_\varphi
\arg\lambda(\varphi+\ii h/2\pi)\,\dd\varphi=-\wind\lambda.
\label{eq:S15}
\end{equation}
The function $\lambda$ is periodic and returns to its initial value when
$\varphi$ completes one period, so the total change of its argument is an
integer multiple of $2\pi$; this integer $\wind\lambda$ is the number of
times $\lambda$ turns counterclockwise around the origin. The acceleration
is therefore the negative of the winding number of the multiplier: as the
modulation phase completes one period, the multiplier, that is, the
complex factor by which a vector in the growing direction is multiplied in
each layer, turns $n$ times clockwise around the origin. The integer that
Avila proved in Eq.~\eqref{eq:S10} thus acquires a concrete meaning in the
uniformly hyperbolic integer phases: it is the negative of the winding
number of the multiplier. A winding number can only be an integer, which
also shows from another angle why the acceleration is an integer.

It remains to show that this result does not depend on the choice of $u$.
The growing direction is a line, and every nonzero multiple of $u$ points
in the same direction, so $u$ is not unique: if $f$ is a periodic,
holomorphic function with no zeros, $fu$ also satisfies the requirements
above. Substituting it into Eq.~\eqref{eq:S12},
$S(\vartheta)f(\vartheta)u(\vartheta)=f(\vartheta)\lambda(\vartheta)u(\vartheta+\alpha)
=\dfrac{f(\vartheta)}{f(\vartheta+\alpha)}\lambda(\vartheta)\cdot f(\vartheta+\alpha)u(\vartheta+\alpha)$,
so with $fu$ the multiplier becomes $\lambda f(\vartheta)/f(\vartheta+\alpha)$.
The extra factor $f(\vartheta)/f(\vartheta+\alpha)$ does not affect the
result: $f(\vartheta)$ and $f(\vartheta+\alpha)$ differ only by a shift of
the phase and have the same winding number, so the winding number of their
ratio is zero; likewise, $\ln|f(\vartheta)|$ and $\ln|f(\vartheta+\alpha)|$
have the same phase average, so the phase average of the log modulus of
the ratio is zero. Therefore Eq.~\eqref{eq:S14} and the winding number do
not depend on the choice of $u$.

\subsection{Winding of the transfer-matrix trace}

The multiplier is not convenient to compute directly. By
Eq.~\eqref{eq:S12}, computing the multiplier at the phase $\vartheta$ first
requires the growing direction $u(\vartheta)$ at that phase. In a
quasicrystal, the growing direction is not an eigenvector of any single
matrix and can only be obtained from propagation: starting from an earlier
phase and propagating a generic state through many layers up to the phase
$\vartheta$, the decaying component is suppressed layer by layer, and the
mixture of the state gradually approaches the growing direction. The exact
growing direction is the limit as the number of propagated layers tends to
infinity. The multiplier can therefore only be obtained approximately,
after a long propagation from an earlier phase. In contrast, the trace of the transfer
matrix needs only the finite product $M_N$ and can be computed directly.
This section shows that the number of times the trace winds around the
origin also gives the acceleration. The product $M_N$ depends on $\varphi$ and $h$ only
through the complex phase $\vartheta$, so, as in the derivation of
Eq.~\eqref{eq:S15}, the derivative of the trace with respect to $h$ equals
its derivative with respect to $\varphi$ multiplied by $\ii/2\pi$. As long
as the trace does not pass through zero as $\varphi$ goes through one
period, taking the logarithm, taking the real part, and averaging over the
phase gives
\begin{equation}
\begin{gathered}
\partial_h\,\frac1N\int_0^1\ln\bigl|\tr M_N(\varphi,h)\bigr|\,\dd\varphi
=-\frac{W_N}{N},\\
W_N=\frac{1}{2\pi}\int_0^1\partial_\varphi\arg\tr M_N(\varphi,h)\,\dd\varphi,
\end{gathered}
\label{eq:S16}
\end{equation}
where the integer $W_N$ is the number of times the trace winds
counterclockwise around the origin as $\varphi$ goes through one period.

We now find $W_N$ in an integer phase with $g_\infty>0$. In this paragraph
we fix $h$, abbreviate $u(\varphi+\ii h/2\pi)$ and
$\lambda(\varphi+\ii h/2\pi)$ as $u(\varphi)$ and $\lambda(\varphi)$, and
take a vector $s(\varphi)$ in the decaying direction. At every phase
$\varphi$, $u(\varphi)$ and $s(\varphi)$ form a basis, and any state can
be decomposed into a growing component along $u(\varphi)$ and a decaying
component along $s(\varphi)$. The trace of a matrix equals the sum of its
diagonal elements in any basis, so, in this basis at the phase $\varphi$,
$\tr M_N(\varphi)$ equals the sum of two terms: the growing component of
$M_Nu(\varphi)$ plus the decaying component of $M_Ns(\varphi)$.

The first term: by Eq.~\eqref{eq:S12},
$M_Nu(\varphi)=\prod_{m=0}^{N-1}\lambda(\varphi+m\alpha)\,u(\varphi+N\alpha)$.
We decompose $u(\varphi+N\alpha)$ at the final phase in the basis at the
initial phase and call its growing component $\Gamma_N(\varphi)$, so the
first term equals $\Gamma_N(\varphi)\prod_m\lambda(\varphi+m\alpha)$.
The quantity $\Gamma_N$ measures how much the growing direction at the final phase
coincides with the growing direction at the initial phase; if
$u(\varphi+N\alpha)$ happens to lie in the decaying direction at the
initial phase, $\Gamma_N$ is zero.

The second term: a vector in the decaying direction is still in the
decaying direction after $N$ layers and is only multiplied by a number.
The relation $\det M_N=1$ means that $M_N$ does not change the area (determinant)
spanned by the two basis vectors; since the growing direction is amplified
by the factor $\prod\lambda$, the decaying direction must be reduced by a
corresponding factor. The angle between the two directions is never zero
on the phase circle, so this reduction factor differs from
$|\prod\lambda|^{-1}$ only by a bounded factor. Together,
\begin{equation}
\begin{gathered}
\tr M_N(\varphi)=\Gamma_N(\varphi)\prod_{m=0}^{N-1}\lambda(\varphi+m\alpha)\\
+O\Bigl(\Bigl|\prod_{m=0}^{N-1}\lambda(\varphi+m\alpha)\Bigr|^{-1}\Bigr),
\end{gathered}
\label{eq:S17}
\end{equation}
uniformly in $\varphi$. By Eq.~\eqref{eq:S11}, the modulus of the product
of multipliers grows at least as a constant times $\ee^{\gamma N}$, so as
long as $\Gamma_N$ is not too small, the second term is much smaller than
the first. If $\Gamma_N$ has no zeros on the phase circle and $N$ is large
enough that the first term exceeds the second everywhere, the winding
number of the trace equals that of the first term. The winding number of a
product is the sum of the winding numbers of its factors, and each factor
$\lambda(\varphi+m\alpha)$ differs from $\lambda$ only by a shift of the
phase, so each has the winding number $\wind\lambda=-n$. Hence
\begin{equation}
W_N=N\wind\lambda+\wind\Gamma_N=-nN+\wind\Gamma_N.
\label{eq:S18}
\end{equation}
This is why the trace carries the bulk integer: the trace is dominated by
the product of multipliers, whose winding number is $-nN$.

\begin{figure}[!tb]
\centering
\includegraphics[width=\linewidth]{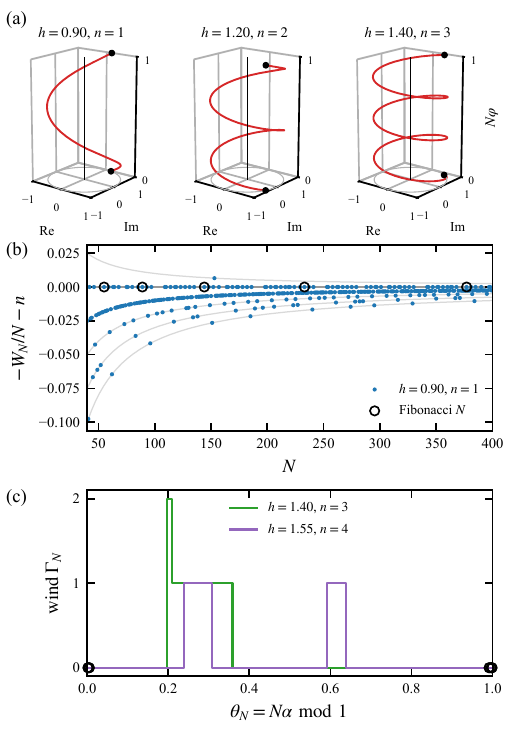}
\caption{Winding of the trace of finite products, $\omega_0\tau=9.03$.
(a)~Normalized trace $\tr M_N/|\tr M_N|$ for $N=144$ over the first $1/N$
of the phase circle, at $h=0.90$, $1.20$, and $1.40$ ($n=1$, 2, and 3).
The two horizontal axes are its real and imaginary parts, and the vertical
axis is $N\varphi$, from 0 to 1; the gray circle is its projection onto the
complex plane, and the vertical line passes through the origin.
(b)~Deviation of the trace winding per layer from the bulk integer,
$-W_N/N-n$, at $h=0.90$ ($n=1$), $40\le N\le400$. The gray curves are
$-\wind\Gamma_N/N$ for the values of $\wind\Gamma_N$ that actually occur,
and open circles mark the Fibonacci lengths.
(c)~End-point winding $\wind\Gamma_N=W_N+nN$ versus
$\theta_N=N\alpha\bmod1$ at $h=1.40$ ($n=3$) and $1.55$ ($n=4$),
$40\le N\le400$; open circles mark the Fibonacci lengths. For $N<40$ the
growing term cannot dominate the trace over the whole phase circle, and
these lengths are not shown.}
\label{fig:S1}
\end{figure}

The second term of Eq.~\eqref{eq:S18} involves only the initial phase
$\varphi$ and the final phase $\varphi+N\alpha$, and not the layers in
between; we call it the end-point term. The end-point term is bounded: $u$
is periodic, so $\Gamma_N$ depends on $N$ only through the fractional part
of $N\alpha$, $\theta_N=N\alpha\bmod1$, and its winding number is a fixed
function of $\theta_N$ that does not depend on $N$. At the four values of
$h$ that we computed ($0.90$, $1.20$, $1.40$, and $1.55$), this function is
a step function, with values between $-3$ and $4$;
Fig.~\ref{fig:S1}(c) shows two of them. Therefore
$\wind\Gamma_N/N\to0$ as $N\to\infty$, and Eq.~\eqref{eq:S18} gives
\begin{equation}
n=\partial_hg_\infty=-\wind\lambda=-\lim_{N\to\infty}\frac{W_N}{N}.
\label{eq:S19}
\end{equation}
At finite $N$, the trace winding per layer, $-W_N/N=n-\wind\Gamma_N/N$, is
in general not an integer; its deviation from $n$ lies on the curves
$-1/N$, $-2/N$, and so on, each curve corresponding to one value of
$\wind\Gamma_N$.

When the final phase comes back close to the initial phase, the end-point
term is exactly zero. If $N\alpha$ is close enough to an integer, then by
periodicity $u(\varphi+N\alpha)$ is close to $u(\varphi)$, and its growing
component $\Gamma_N$ is close to 1. As long as the distance from
$\Gamma_N$ to 1 is smaller than 1 over the whole phase circle, the curve
$\Gamma_N(\varphi)$ cannot wind around the origin, and so
$\wind\Gamma_N=0$. For the golden mean, $N\alpha$ comes closest to an
integer at the Fibonacci numbers, where its distance to the nearest
integer is smaller than $1/N$.

Figure~\ref{fig:S1} illustrates these conclusions. Figure~\ref{fig:S1}(a)
takes $N=144$ and three values of $h$: as $\varphi$ goes through $[0,1]$,
the trace turns $nN$ times clockwise around the origin at an almost
constant rate, with $-\arg\tr M_N/(2\pi N)$ differing from $n\varphi$ by
less than $3\times10^{-4}$, so within the first $1/N$ of the phase circle
the trace turns close to $n$ times. Figure~\ref{fig:S1}(b) takes $h=0.90$
($n=1$): the deviation $-W_N/N-n$ lies on a few curves $-\wind\Gamma_N/N$
and is zero at the Fibonacci lengths. Figure~\ref{fig:S1}(c) takes
$h=1.40$ and $1.55$ ($n=3$ and $4$): the end-point windings of all lengths
lie on the same step function of $\theta_N$, which is zero when $\theta_N$
is close to 0 and 1.

We also took one $h$ in each of the integer phases $n=1$--$4$ ($0.90$,
$1.20$, $1.40$, and $1.55$) and checked the end-point term at several
lengths. When $\theta_N$ is close to 0 or 1, the end-point term is zero
and $-W_N/N$ equals $n$ exactly: for example, at the Fibonacci length
$N=144$ this holds at all four values of $h$. When $\theta_N$ is far from
0 and 1, the end-point term can be nonzero, and whether it is zero also
depends on $h$: for example, for $N=200$, $N\alpha$ is $0.39$ away from the
nearest integer, and at $h=0.90$, $\wind\Gamma_N=1$ and
$-W_N/N=1-1/200=0.995$.

\section{Phase boundaries}

\subsection{Fourier coefficients of rational approximants}

The integer phases are separated by corners of $g_\infty$, where the slope
jumps by an integer. A growth rate computed with a finite number of layers
has no corners: its slope changes gradually near a corner of $g_\infty$,
so the exact position of the corner is hard to read from it. We instead
locate the corners with rational approximations of the irrational rotation
number $\alpha$. The best rational approximations of $\alpha$ are given by
successive truncations of its continued fraction and are called the
convergents $p/q$; for example, the convergents of the golden mean are
$\frac12,\frac23,\frac35,\frac58,\ldots$, whose denominators are Fibonacci
numbers. Replacing $\alpha$ by $p/q$ turns the quasiperiodic sequence into
a periodic sequence that repeats every $q$ layers. We call this periodic
sequence a rational approximant and one of its periods ($q$ layers) a
cell. The transfer matrix of the cell is
\begin{equation}
M_q^{(p/q)}(\varphi,h)=S_h\!\left(\varphi+\frac{(q-1)p}{q}\right)\cdots S_h(\varphi).
\label{eq:S20}
\end{equation}
The layer matrix depends on $\varphi$ through $\sin(2\pi\varphi+\ii h)$;
when $\varphi$ increases by 1, the modulation phase increases by $2\pi$ and
the layer matrix does not change. Shifting $\varphi$ by $p/q$ turns the
$m$th factor into the original $(m+1)$th factor, and the last factor
becomes $S_h(\varphi+p)=S_h(\varphi)$, the original first factor; so the
shift only reorders the $q$ factors cyclically, and the trace of a matrix
product does not change under a cyclic reordering. Since $p$ and $q$ are
coprime, integer combinations of $p/q$ and 1 give every integer multiple of
$1/q$, so the trace, as a function of the initial phase $\varphi$, has the
period $1/q$. This is a period in the initial phase, not the same thing as
the period in time, ``repeating every $q$ layers''. Since the trace is a
periodic function of $\varphi$, it can be expanded in a Fourier series:
\begin{equation}
\begin{gathered}
\tr M_q^{(p/q)}(\varphi,h)=\sum_{n\in\mathbb Z} C_n^{(q)}(h)\,\ee^{-2\pi\ii nq\varphi},\\
C_n^{(q)}(h)=q\int_0^{1/q}\tr M_q^{(p/q)}(\varphi,h)\,\ee^{2\pi\ii nq\varphi}\,\dd\varphi.
\end{gathered}
\label{eq:S21}
\end{equation}

\begin{figure}[!tb]
\centering
\includegraphics[width=\linewidth]{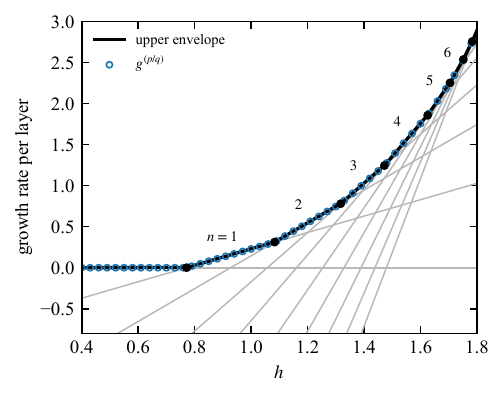}
\caption{Fourier-coefficient lines of the cell $p/q=89/144$ at
$\omega_0\tau=9.03$. Gray lines: the lines
$q^{-1}\ln|C_n^{(q)}(h)|=b_n^{(q)}+nh$ of Eq.~\eqref{eq:S23} for
$n=0,\ldots,9$, and the zero line; the line of $n=0$ lies slightly below
zero. Black line: the upper envelope, that is, the right side of
Eq.~\eqref{eq:S25} without the error term. Circles: the phase-averaged
growth rate $g^{(p/q)}$ computed directly from the cell
[Eq.~\eqref{eq:S24}]. Black dots: corners of the envelope, at $h=0.772$
(where the line of $n=1$ crosses zero) and at the crossings
$h_{c,n}^{(q)}$ of Eq.~\eqref{eq:S26}. Numbers give the order of the top
line.}
\label{fig:S2}
\end{figure}

We now look at how the Fourier coefficients change with $h$. The key point
is that $h$ does not appear in the trace on its own: the trace depends on
$\varphi$ and $h$ only through the combination $\vartheta=\varphi+\ii h/2\pi$
(Sec.~S1\,B). The trace at $h$ is therefore the trace at $h=0$ with
$\varphi$ replaced by $\varphi+\ii h/2\pi$. Inside the analytic strip the
trace is an analytic periodic function of $\vartheta$, and its Fourier
series converges everywhere inside the strip, so the replacement can be
made in each term of the series separately. After the replacement, the
$n$th term $C_n^{(q)}(0)\ee^{-2\pi\ii nq\varphi}$ becomes
\begin{equation}
C_n^{(q)}(0)\,\ee^{-2\pi\ii nq(\varphi+\ii h/2\pi)}
=C_n^{(q)}(0)\,\ee^{nqh}\,\ee^{-2\pi\ii nq\varphi}.
\label{eq:S22}
\end{equation}
Comparing term by term with the series~\eqref{eq:S21} at $h$ gives
$C_n^{(q)}(h)=C_n^{(q)}(0)\,\ee^{nqh}$. After taking the logarithm,
$q^{-1}\ln|C_n^{(q)}(h)|$ is a linear function of $h$ with the integer
slope $n$. We call its graph the line of order $n$ [gray lines in
Fig.~\ref{fig:S2}]. Numerically, it is more convenient to fix this line
with the coefficient at a reference value $h_0$ inside the strip:
\begin{equation}
\begin{aligned}
C_n^{(q)}(h)&=C_n^{(q)}(h_0)\,\ee^{nq(h-h_0)},\\
\tfrac1q\ln|C_n^{(q)}(h)|&=b_n^{(q)}+nh,\\
b_n^{(q)}&=\tfrac1q\ln|C_n^{(q)}(h_0)|-nh_0.
\end{aligned}
\label{eq:S23}
\end{equation}
The intercept $b_n^{(q)}$ does not depend on the choice of $h_0$ and is
the intercept of the main text; Figs.~3(a) and 3(b) of the main text show
these coefficients.

The integer slope of each line has a simple origin, which can be phrased as the action of an imaginary gauge field. Because the layer matrix depends on $\varphi$ and $h$ only through $\vartheta$, it can be expanded in harmonics of the modulation phase,
\begin{equation}
\begin{aligned}
S_h(\varphi)&=\sum_{k\in\mathbb Z}s_k\,e^{-2\pi ik(\varphi+ih/2\pi)}\\
&=\sum_{k\in\mathbb Z}s_k\,e^{kh}\,e^{-2\pi ik\varphi},
\end{aligned}
\label{eq:gauge-harmonics}
\end{equation}
where the $2\times2$ matrices $s_k$ are the Fourier coefficients of the layer matrix at $h=0$. Complexification therefore multiplies the harmonic of order $k$ by $e^{kh}$. When the $q$ factors of Eq.~\eqref{eq:S20} are multiplied out, each term takes one harmonic from each factor, of orders $k_1,\dots,k_q$, and contributes to the trace a harmonic of order $k_1+\dots+k_q$ with the weight $e^{(k_1+\dots+k_q)h}$. The coefficient $C_n^{(q)}$ collects all terms with $k_1+\dots+k_q=nq$, that is, with $n$ net harmonic steps per layer, and every such term carries the same factor $e^{nqh}$, in agreement with Eq.~\eqref{eq:S22}. If the harmonic orders are regarded as the sites of a lattice, each factor shifts the order by $k$ sites, and $e^{kh}$ is exactly the weight that an imaginary vector potential attaches to a hop of $k$ sites in the Hatano--Nelson model~\cite{HatanoNelson1996}; $h$ plays the role of this imaginary gauge field. In the non-Hermitian AAH model the layer matrix contains only the harmonics $k=0,\pm1$, so each layer shifts the order by at most one site and the slope is at most one. The layer matrix here contains harmonics of all orders, so this counting gives no bound; which order dominates is decided by the magnitudes of the coefficients, and near the edge of the strip the high orders become large (Sec.~S3\,B).

These lines determine the growth rate of the rational approximant. A
matrix in $\mathrm{SL}(2,\mathbb C)$ has two eigenvalues $\mu$ and
$1/\mu$, and $\mu+1/\mu$ equals the trace. Therefore, when the trace is
large, the spectral radius $\rho$ of the matrix (the largest modulus of
its eigenvalues) is close to the modulus of the trace. Where one term of
Eq.~\eqref{eq:S21} dominates and is large, the growth rate per layer of
the rational approximant, $q^{-1}\ln\rho\,M_q^{(p/q)}$, is close to the
line of this term; the largest term corresponds to the highest line.
Since $\rho\ge1$, the growth rate is never negative. We call the function
of $h$ obtained by taking, at each $h$, the maximum of zero and the lines
of all orders the upper envelope of these lines [black line in
Fig.~\ref{fig:S2}]; where the envelope is positive, the line that forms the
envelope is called the top line. For the phase-averaged growth rate per
layer
\begin{equation}
g^{(p/q)}(h)=\frac1q\int_0^1\ln\rho\bigl(M_q^{(p/q)}(\varphi,h)\bigr)\,\dd\varphi,
\label{eq:S24}
\end{equation}
this correspondence does not require that one term dominates: on any
closed interval inside the strip, the difference between $g^{(p/q)}$ and
the upper envelope tends to zero uniformly as
$q\to\infty$~\cite{Avila2015},
\begin{equation}
g^{(p/q)}(h)=\max\Bigl\{0,\;\max_{n}\bigl[b_n^{(q)}+nh\bigr]\Bigr\}+o(1),
\qquad q\to\infty.
\label{eq:S25}
\end{equation}
For the cell $p/q=89/144$ and $0.4\le h\le1.8$, the directly computed
$g^{(p/q)}$ differs from the envelope by less than $2\times10^{-3}$
[circles in Fig.~\ref{fig:S2}]. Where all lines lie below zero, for
example for $h<0.77$, the envelope and the growth rate are both zero.

The growth rate $g^{(p/q)}$ is the Lyapunov exponent of the cocycle with the rotation
number $p/q$~\cite{Avila2015}, and $g_\infty$ is the Lyapunov exponent
with the rotation number $\alpha$. The Lyapunov exponent of an analytic
$\mathrm{SL}(2,\mathbb C)$ cocycle is continuous in the rotation number at
irrational rotation numbers~\cite{JKS2009}, so along the convergents
$g^{(p/q)}$ tends to $g_\infty$ at every $h$; by Eq.~\eqref{eq:S25}, the
envelope also tends to $g_\infty$. The envelope and $g_\infty$ are both
convex functions, and a convex function cannot oscillate back and forth,
so when the values approach each other, the slopes also approach each
other: the slope of the envelope, that is, the order of the top line,
tends to the slope of $g_\infty$. In an integer phase with $g_\infty>0$,
this order is an integer that tends to $n$, so it equals the bulk integer
$n$ once $q$ is large enough.

Thus the corners of $g_\infty$ are the limits of the corners of the
envelope [black dots in Fig.~\ref{fig:S2}]. The envelope has corners at
two kinds of places: where the top line changes to another line, and
where the top line crosses zero. The lines of orders $n$ and $n+1$ cross
at
\begin{equation}
h_{c,n}^{(q)}=b_n^{(q)}-b_{n+1}^{(q)}
=h_0+\frac1q\ln\left|\frac{C_n^{(q)}(h_0)}{C_{n+1}^{(q)}(h_0)}\right|.
\label{eq:S26}
\end{equation}
When the top line changes from order $n$ to order $n+1$, the crossing
$h_{c,n}^{(q)}$ approximates the boundary between the integer phases $n$
and $n+1$. Crossings between lines below the envelope do not correspond to
any corner. Equation~\eqref{eq:S26} needs only two Fourier coefficients at
the same $h_0$ to fix a boundary; neither the growth rate nor its slope
needs to be computed.

As $q$ increases, the crossings converge: the first five crossings change
by less than $10^{-5}$ between $q=987$ and $q=1597$. For the largest
rational approximant we computed, $q=1597$, the crossings are
$h_{c,n}^{(1597)}=1.0832$, $1.3174$, $1.4723$, $1.6253$, and $1.7043$
($n=1,\ldots,5$), and they differ from the boundaries obtained by direct
propagation of the irrational sequence by less than $2\times10^{-4}$
(Sec.~S6).

\subsection{Edge of the analytic strip}

This section shows that the growth rate remains finite at the edge of the
strip and that the boundaries between integer phases accumulate without
limit at the edge. The edge is special because the permittivity
of a few temporal layers tends to zero. The modulation phase of temporal
layer $m$ (Sec.~S1\,A) is $X_m$; as $h\to h_\star$, both the real and the
imaginary parts of the permittivity tend to zero for the temporal layers
whose $X_m$ is close to $3\pi/2$. At $X_m=3\pi/2$ the permittivity is the
real number $\varepsilon_R-\delta\varepsilon\cosh h$, which is exactly zero
at $h=h_\star$. To treat these temporal layers, we regard the modulation
phase as a continuous variable $X$ and write
\begin{equation}
\begin{gathered}
\varepsilon(X,h)=\varepsilon_R+\delta\varepsilon\sin(X+\ii h),\\
z(X,h)=\sqrt{\varepsilon_R/\varepsilon(X,h)},
\end{gathered}
\label{eq:S27}
\end{equation}
where we take the branch of the square root with positive real part;
inside the strip $\operatorname{Re}\varepsilon>0$, so this branch always
exists. The $z$ here is the $z_m$ of Sec.~S1\,A written as a continuous
function of the phase: the $z_m$ of layer $m$ equals $z(X_m,h)$.

We first give an upper bound on $g_\infty$. The norm of a product of
matrices does not exceed the product of the norms of the layer matrices,
so $\ln\|M_N\|$ does not exceed the sum of $\ln\|S_h\|$ over the layers.
We then average over the initial phase: a shift of the phase by $\alpha$
does not change the average, so every layer contributes the same amount to
the average, namely the phase average of $\ln\|S_h\|$ for a single layer.
Therefore an estimate of the norm of a single layer matrix gives an upper
bound on $g_\infty$.

The norm of a single layer matrix can be estimated directly from its
matrix elements. By Eq.~\eqref{eq:S4}, the layer matrix of a temporal
layer with the modulation phase $X$ is
\[
\begin{gathered}
S_h=\begin{pmatrix}\cos(\omega_0\tau z)&-\ii\sin(\omega_0\tau z)/z\\
-\ii z\sin(\omega_0\tau z)&\cos(\omega_0\tau z)\end{pmatrix},\\ z=z(X,h).
\end{gathered}
\]
The spectral norm of a matrix does not exceed its Frobenius norm, the
square root of the sum of the squared moduli of its
elements~\cite{HornJohnson2013}. For any complex number $w$, $|\cos w|$
and $|\sin w|$ do not exceed $\cosh(\operatorname{Im}w)$ and hence do not
exceed $\ee^{|\operatorname{Im}w|}$. Taking $w=\omega_0\tau z$, the moduli
of the four matrix elements do not exceed
$\ee^{\omega_0\tau|\operatorname{Im}z|}$ times $1$, $|z|^{-1}$, $|z|$,
and $1$, respectively, so
\[
\begin{gathered}
\|S_h\|\le\ee^{\omega_0\tau|\operatorname{Im}z|}\bigl(2+|z|^2+|z|^{-2}\bigr)^{1/2}\\
=\ee^{\omega_0\tau|\operatorname{Im}z|}\bigl(|z|+|z|^{-1}\bigr),
\end{gathered}
\]
that is,
\begin{equation}
\ln\|S_h\|\le\omega_0\tau|\operatorname{Im}z|+\ln\bigl(|z|+|z|^{-1}\bigr).
\label{eq:S28}
\end{equation}
The first term on the right is the growth exponent of the faster-growing
of the two waves in the layer: over one layer, the two waves
$D\propto\ee^{\mp\ii zs}$ in the layer are multiplied by
$\ee^{\mp\ii\omega_0\tau z}$, and the larger modulus is
$\ee^{\omega_0\tau|\operatorname{Im}z|}$. The second term comes from the
factors $z$ and $1/z$ in the off-diagonal elements and becomes large when
$|z|$ is far from 1, that is, when the permittivity is far from
$\varepsilon_R$. Inserting Eq.~\eqref{eq:S28} into the phase average above
and changing the variable to $X=2\pi\varphi$ gives, for any $\alpha$,
\begin{equation}
\begin{gathered}
0\le g_\infty(h)\le U(h)\equiv\frac{1}{2\pi}\int_0^{2\pi}\\
\Bigl[\omega_0\tau|\operatorname{Im}z(X,h)|
\\ +\ln\bigl(|z(X,h)|+|z(X,h)|^{-1}\bigr)\Bigr]\dd X.
\end{gathered}
\label{eq:S29}
\end{equation}
The lower bound comes from $\det M_N=1$: the two singular values of $M_N$
have product 1, so the larger one, $\|M_N\|$, is not smaller than 1.
Equation~\eqref{eq:S29} shows that, for every $h$ inside the strip, the
growth rate does not exceed a quantity $U(h)$ that is set by the
properties of a single layer alone and does not depend on the order of the
layers. The next paragraph shows that $U$ is still finite at the edge of
the strip, so that $g_\infty$ is also finite at the edge.

The upper bound $U$ remains finite at the edge of the strip, although the
integrand diverges near the zero of the permittivity. Write
$d=h_\star-h>0$ and $\Delta X=X-3\pi/2$. Near the zero,
\begin{equation}
\begin{gathered}
\varepsilon(3\pi/2+\Delta X,h_\star-d)\\
=(\varepsilon_R^2-\delta\varepsilon^2)^{1/2}(d+\ii\Delta X)+O(d^2+\Delta X^2),
\end{gathered}
\label{eq:S30}
\end{equation}
so in a neighborhood of $(d,\Delta X)=(0,0)$,
$|\varepsilon|\ge c_\varepsilon(d^2+\Delta X^2)^{1/2}$ with a constant
$c_\varepsilon>0$. Hence $|z|$ does not exceed a constant times
$|\Delta X|^{-1/2}$, and in this neighborhood the integrand of
Eq.~\eqref{eq:S29} does not exceed a constant times
$1+|\Delta X|^{-1/2}+|\ln|\Delta X||$ for all $d$; away from $\Delta X=0$
the integrand is bounded anyway. Both $|\Delta X|^{-1/2}$ and
$|\ln|\Delta X||$ are integrable, so the integrand is bounded by an
integrable function that does not depend on $d$, and we may let $d\to0$
before integrating (dominated convergence theorem), which gives
$U_\star=\lim_{h\uparrow h_\star}U(h)<\infty$. Moreover, by Sec.~S2\,A the
slope of $g_\infty$ is nonnegative for $h>0$, so $g_\infty$ does not
decrease with $h$, and
\begin{equation}
g_\star\equiv\lim_{h\uparrow h_\star}g_\infty(h)
=\sup_{0\le h<h_\star}g_\infty(h)\le U_\star<\infty.
\label{eq:S31}
\end{equation}
The bound at the edge is $U_\star=5.886$, and the numerically computed edge
value is $g_\star=4.566$ (Sec.~S6).

We now turn to the second conclusion: the phase boundaries accumulate at the
edge of the strip. The growth rate is finite at the edge, but its slope,
the integer $n$, can diverge there; once the slope diverges, the
boundaries between adjacent integer phases must crowd closer and closer
together. The key is therefore to find how the slope diverges at the edge.
To do so, we split the growth rate into two parts. The first part, $J(h)$,
is the phase average of the growth exponent
$\omega_0\tau|\operatorname{Im}z|$ of the faster-growing wave in a single
temporal layer [the first term of Eq.~\eqref{eq:S28}]; it can be computed
exactly, and the singularity at the edge comes from this part. The second
part, $R(h)$, is the rest of the growth rate, which depends on the order
of the temporal layers; below we will assume that it changes slowly at the
edge. That is,
\begin{equation}
\begin{gathered}
J(h)=\frac{\omega_0\tau}{2\pi}\int_0^{2\pi}|\operatorname{Im}z(X,h)|\,\dd X,\\ R(h)=g_\infty(h)-J(h).
\end{gathered}
\label{eq:S32}
\end{equation}
The derivative of $J$ can be computed exactly. For $0<h<h_\star$,
$\operatorname{Im}\varepsilon=\delta\varepsilon\sinh h\cos X$, so the sign
of $\operatorname{Im}z$ is opposite to that of $\cos X$, and the sign
changes are fixed at $X=\pi/2$ and $3\pi/2$ and do not move with $h$.
The quantity $z$ is a holomorphic function of $X+\ii h$, so by the Cauchy--Riemann
relations $\partial_h\operatorname{Im}z=\partial_X\operatorname{Re}z$. On
each interval of constant sign, therefore,
$\partial_h|\operatorname{Im}z|=\pm\partial_X\operatorname{Re}z$, and after
integration over $X$ only the values of $\operatorname{Re}z$ at the ends
of the interval remain. At both ends $z$ is real: at $X=\pi/2$,
$z=[\varepsilon_R/(\varepsilon_R+\delta\varepsilon\cosh h)]^{1/2}$, and at
$X=3\pi/2$, $z=[\varepsilon_R/(\varepsilon_R-\delta\varepsilon\cosh h)]^{1/2}$.
The result is
\begin{equation}
\begin{gathered}
J'(h)=\frac{\omega_0\tau\sqrt{\varepsilon_R}}{\pi}\\
\left[\frac{1}{\sqrt{\varepsilon_R-\delta\varepsilon\cosh h}}
-\frac{1}{\sqrt{\varepsilon_R+\delta\varepsilon\cosh h}}\right],\\ J(0)=0.
\end{gathered}
\label{eq:S33}
\end{equation}

The first term of Eq.~\eqref{eq:S33} diverges at the edge; it comes from
the temporal layers near $X=3\pi/2$. The permittivity of these layers is
close to $\varepsilon_{\min}=\varepsilon_R-\delta\varepsilon\cosh h$, and
near the edge
$\varepsilon_{\min}=(\varepsilon_R^2-\delta\varepsilon^2)^{1/2}d+O(d^2)$
tends to zero. The phase that the wave accumulates within one such layer,
$\omega_0\tau\sqrt{\varepsilon_R/\varepsilon_{\min}}$, therefore diverges
as $d^{-1/2}$. We denote the coefficient of this divergence by $A_\star$:
\begin{equation}
\begin{gathered}
\omega_0\tau\sqrt{\varepsilon_R/\varepsilon_{\min}}\\
=\frac{\pi A_\star}{\sqrt d}\bigl[1+O(d)\bigr],\\
A_\star=\frac{\omega_0\tau\sqrt{\varepsilon_R}}
{\pi(\varepsilon_R^2-\delta\varepsilon^2)^{1/4}}.
\end{gathered}
\label{eq:S34}
\end{equation}
That is, $A_\star/\sqrt d$ is the number of half periods that the wave goes
through within the temporal layer of smallest permittivity. The coefficient $A_\star$ is
set by $\omega_0\tau$, $\varepsilon_R$, and $\delta\varepsilon$ alone and
contains no adjustable parameter. Writing the first term of
Eq.~\eqref{eq:S33} with it, $J'=A_\star/\sqrt d+O(1)$, and integrating over
$h$ gives
\begin{equation}
J_\star-J(h)=2A_\star\sqrt d+O(d),
\label{eq:S35}
\end{equation}
where $J_\star=\lim_{h\uparrow h_\star}J(h)$, and $J_\star=5.106$.

\begin{figure}[!tb]
\centering
\includegraphics[width=\linewidth]{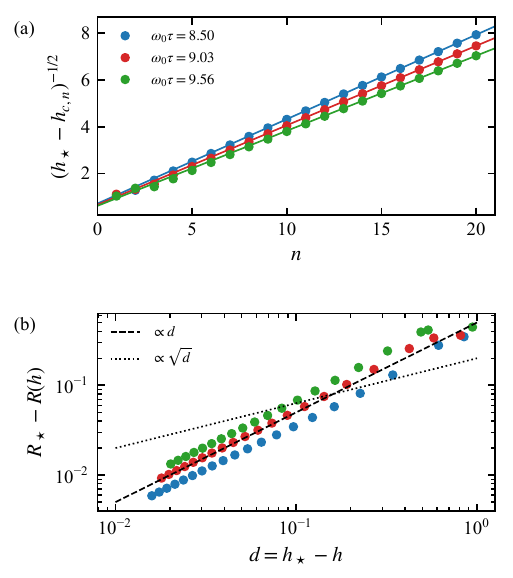}
\caption{Accumulation of the integer phases at the edge of the analytic
strip. (a)~Inverse square root of the distance from the $n\to n+1$
boundary to the edge, $(h_\star-h_{c,n})^{-1/2}$, computed from
Eq.~\eqref{eq:S26} at $q=2584$ for three values of $\omega_0\tau$. The
straight lines have the slope $1/A_\star$ of Eq.~\eqref{eq:S39} and pass
through the last point; at finite $n$ the data agree with
$(n+n_0)/A_\star$, $n_0\simeq2$. (b)~Approach of the remainder $R$
[Eq.~\eqref{eq:S32}] to its edge value at the corners of $g_\infty$.
The edge value $R_\star$ is obtained by linear extrapolation and, for
$\omega_0\tau=9.03$, agrees with the independently computed
$g_\star-J_\star$. Dashed and dotted lines indicate linear and square-root
scaling, respectively.}
\label{fig:S3}
\end{figure}

The remainder $R$ cannot be computed exactly. We assume that it has no
singularity of the $\sqrt d$ type at the edge and instead tends linearly
to its edge value $R_\star=g_\star-J_\star$:
\begin{equation}
R_\star-R(h)=\kappa\,d+o(d),\qquad h\uparrow h_\star.
\label{eq:S36}
\end{equation}
This assumption has numerical support. We compute $R$ at the corners of
$g_\infty$, because at a corner the growth rate is the common value of two
adjacent lines [Eq.~\eqref{eq:S23}] and can be read directly from the
Fourier coefficients; the coefficients are taken at $q=2584$, and changing
to $q=1597$ changes the result by less than $1.3\%$ of $\kappa d$. At the
corners, $R_\star-R$ lies on a straight line as a function of $d$, with
the slopes $\kappa\simeq0.37$, $0.52$, and $0.66$ for $\omega_0\tau=8.50$,
$9.03$, and $9.56$, respectively [Fig.~\ref{fig:S3}(b)]; for
$\omega_0\tau=9.03$ the extrapolated $R_\star=-0.540$ agrees with the
independently computed $g_\star-J_\star$. Under this assumption the
$\sqrt d$ singularity of the growth rate comes entirely from $J$, and
adding Eqs.~\eqref{eq:S35} and~\eqref{eq:S36} gives
\begin{equation}
g_\star-g_\infty(h)=2A_\star\sqrt d+O(d).
\label{eq:S37}
\end{equation}

The slope follows from Eq.~\eqref{eq:S37}. The bulk growth rate $g_\infty$ is a convex
function, so its right derivative at $h$ lies between the slopes of two
secants: one over the interval $[h-\delta d,h]$ and the other over
$[h,h+\delta d]$, where $0<\delta<1$. Inserting Eq.~\eqref{eq:S37}, the
slopes of the two secants multiplied by $\sqrt d$ tend, at fixed $\delta$
and $d\to0$, to $2A_\star(\sqrt{1+\delta}-1)/\delta$ and
$2A_\star(1-\sqrt{1-\delta})/\delta$, respectively; letting then
$\delta\to0$, both tend to $A_\star$. Therefore
\begin{equation}
n(h)=\partial_h^+g_\infty(h)=\frac{A_\star}{\sqrt d}\bigl[1+o(1)\bigr].
\label{eq:S38}
\end{equation}
Compared with Eq.~\eqref{eq:S34}, this shows that, near the edge, the
integer $n$ is approximately the number of half periods that the wave goes
through within the temporal layer of smallest permittivity, and that it
diverges as $d\to0$. At the boundary $h_{c,n}$ between the integer phases
$n$ and $n+1$, the right derivative equals $n+1$; applying
Eq.~\eqref{eq:S38} at this point and noting that the ratio of $n+1$ to $n$
tends to 1 as $n\to\infty$, we obtain the accumulation law of the phase
boundaries,
\begin{equation}
\bigl(h_\star-h_{c,n}\bigr)^{-1/2}=\frac{n}{A_\star}\bigl[1+o(1)\bigr],
\qquad n\to\infty.
\label{eq:S39}
\end{equation}

The inverse square root of the distance from a boundary to the edge grows
linearly with the integer, at the rate $1/A_\star$, which contains no
adjustable parameter. Figure~\ref{fig:S3}(a) tests this law with the
boundaries computed from Eq.~\eqref{eq:S26} at $q=2584$: at all three
frequencies, for larger $n$, the difference of $(h_\star-h_{c,n})^{-1/2}$
between adjacent boundaries differs from $1/A_\star$ by less than $0.5\%$.

In summary, at the edge of the analytic strip the growth rate tends to the
finite value $g_\star$ [Eq.~\eqref{eq:S31}], while the integer $n$ diverges
following $A_\star/\sqrt d$, the phase boundaries accumulate at the edge
following Eq.~\eqref{eq:S39}, and the widths of the integer phases tend to
zero.

\section{Conditions for the integer response}

The derivation in Sec.~S2 uses only two properties of the modulation: the
layer matrix is an analytic function of the complex phase $\vartheta$,
and the rotation number is irrational. As long as a perturbation preserves
these two properties, the slope of the growth rate is still an integer;
moreover, under a continuous deformation, this integer does not change as
long as no phase boundary crosses the $h$ under consideration.
Section~S4\,A tests two kinds of perturbations that preserve these two
properties: distortions of the modulation waveform and local defects.
Section~S4\,B changes the order of the layers and compares the
quasiperiodic sequence with periodic and random sequences.

\subsection{Waveform distortion and local defects}

We first ask whether a distortion of the modulation waveform changes the
integer. Adding a second harmonic $\eta\,\delta\varepsilon\sin(2X+\psi)$
to the real waveform and, as for the fundamental, replacing $X$ by
$X+\ii h$ gives
\begin{equation}
\begin{gathered}
\varepsilon_m^{(\eta)}=\varepsilon_R+\delta\varepsilon\sin Z_m
+\eta\,\delta\varepsilon\sin(2Z_m+\psi),\\ Z_m=X_m+\ii h,
\end{gathered}
\label{eq:S40}
\end{equation}
where $\eta$ is the relative amplitude and $\psi$ is the phase difference.
After the continuation, the real and imaginary parts of the second
harmonic have the amplitudes $\eta\cosh2h$ and $\eta\sinh2h$ relative to
$\delta\varepsilon$. The layer matrix still depends on $\varphi$ and $h$
only through $\vartheta$, and this is exactly the property used in deriving
Eqs.~\eqref{eq:S10} and~\eqref{eq:S15}. The distorted medium therefore
still satisfies the conditions of Avila's theorem, and the acceleration is
still an integer. The same conclusion holds for any harmonic, as long as
$X$ in it is likewise replaced by $X+\ii h$. Figure~3(e) of the main text
takes $\eta=0.04$ and $\psi=0.37$; the permittivity then first vanishes at
$|h|=1.7614$, and the analytic strip becomes $|h|<1.7614$, narrower than
for $\eta=0$. The distortion moves the boundaries between the integer
phases and changes the intercepts of the straight segments, but it does
not change the integer that belongs to each integer phase. The reason is
that, as $\eta$ changes continuously, the multiplier of Eq.~\eqref{eq:S12}
changes continuously as long as the $h$ under consideration stays in the
same integer phase, that is, as long as the permittivity never vanishes
and the cocycle stays uniformly hyperbolic. A winding number is an integer
and cannot jump under a continuous change, so by Eq.~\eqref{eq:S15} the
slope stays the same and only the intercept changes:
\begin{equation}
g_{\infty,\eta}(h)=b(\eta)+nh,\qquad \partial_\eta\partial_hg_{\infty,\eta}=0.
\label{eq:S41}
\end{equation}
In Fig.~3(e) of the main text, the $n=2$ plateau that all values of $\eta$
share at $h=1.20$ shows this.

\begin{figure}[!tb]
\centering
\includegraphics[width=\linewidth]{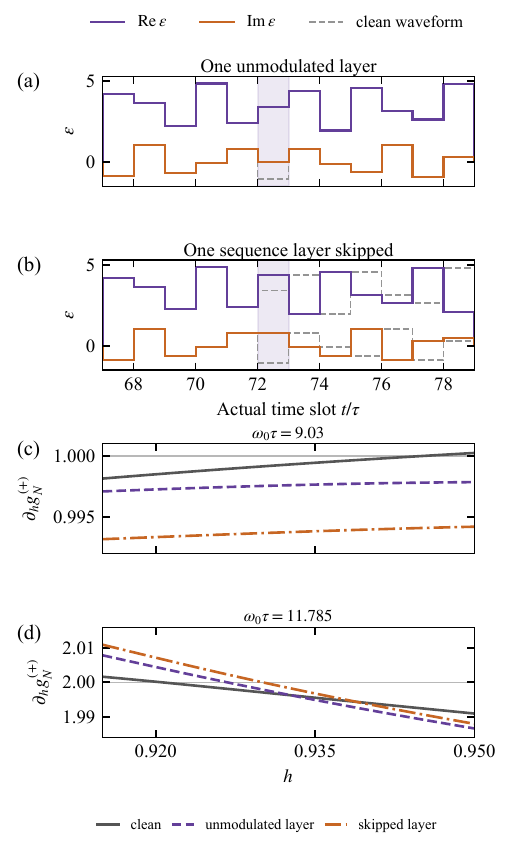}
\caption{Local temporal defects in the golden-mean sequence,
$\varphi_0=0$. (a)~The permittivity of layer $j=72$ is set to
$\varepsilon_R$. (b)~Layer $j=72$ of the sequence is skipped; the
subsequent layers take the next indices of the sequence in turn, and one
layer is added at the end so that the total duration remains $N\tau$. Both
waveforms are drawn at $h=0.925$; the gray dashed lines are the waveforms
without defects, and the shading marks the first affected temporal layer.
(c),(d)~Forward-input response $s_N^{(+)}$ for $N=512$ at
$\omega_0\tau=9.03$ and $11.785$, where the integers without defects are
1 and 2, respectively.}
\label{fig:S4}
\end{figure}

We next ask whether a local defect that affects only one temporal layer
changes the integer. We replace the layer matrix $S_j$ of layer $j$ by any
invertible matrix $\widetilde S_j$ and write the product of the layers
before the defect as $\mathcal Q_j=S_{j-1}\cdots S_0$. The product with
the defect can be written as the original product multiplied on the right
by a fixed matrix:
\begin{equation}
\begin{gathered}
\widetilde M_N=M_N\mathcal K_j,\\
\mathcal K_j=\mathcal Q_j^{-1}S_j^{-1}\widetilde S_j\mathcal Q_j,\\
\|\mathcal K_j^{-1}\|^{-1}\le\frac{\|\widetilde M_N\|}{\|M_N\|}\le\|\mathcal K_j\|.
\end{gathered}
\label{eq:S42}
\end{equation}
The matrix $\mathcal K_j$ does not depend on $N$, so the ratio of the norms with and
without the defect is bounded by two constants that do not depend on $N$;
after taking the logarithm and dividing by $N$, the contribution of this
ratio tends to zero, so the bulk growth rate and its slope do not change,
and a local defect does not change the integer. The same conclusion holds
for defects in any finite number of temporal layers. Another kind of
defect skips layer $j$: the subsequent layers move forward by one index in
turn, and one layer is added at the end so that the total duration remains
$N\tau$. The product is then the original product multiplied on the left
by the added last layer and on the right by a fixed matrix that does not
depend on $N$; the norm of the added last layer is bounded, so the
conclusion again does not change.

Figure~\ref{fig:S4} shows these two kinds of defects on a single finite
sequence. We plot the forward-input response $s_N^{(+)}$: the initial
phase is fixed, the input is the forward wave
$v_{\rm in}=(1,1)^{\mathsf T}$, the growth rate is taken as
$N^{-1}\ln(\|M_Nv_{\rm in}\|/\|v_{\rm in}\|)$, and it is then
differentiated with respect to $h$; there is no average over the initial
phase. The computed responses are all close to the integer of the
sequence without defects. Small deviations from the integer remain
because $N$ is finite: the contribution of the bounded factor in
Eq.~\eqref{eq:S42} to the growth rate decreases as $1/N$ and vanishes only
as $N\to\infty$.

\subsection{Periodic and random sequences}

This section shows that the integer plateaus are a property specific to
quasiperiodic sequences: if the single-layer permittivity and its
dependence on $h$ are kept unchanged and only the quasiperiodic order is
replaced by a periodic or random order, a single sequence no longer gives
integer plateaus. Consider first the periodic order. When the rotation
number is a rational number $p/q$, the sequence repeats every $q$ layers,
and a single sequence is the repetition of the cell
$M_q^{(p/q)}(\varphi_0,h)$ of Sec.~S3\,A. When the cell has an eigenvalue
$\mu$ with modulus larger than 1, the growth rate per layer of the
infinitely repeated cell is $\ln|\mu|/q$. Differentiating
$\mu+1/\mu=\tr M_q^{(p/q)}$ gives its response to $h$, which we call the
Floquet response:
\begin{equation}
\begin{gathered}
g_{\rm per}(\varphi_0,h)=\frac{\ln|\mu|}{q},\\
s_{\rm per}=\partial_hg_{\rm per}
=\frac1q\operatorname{Re}\frac{\partial_h\tr M_q^{(p/q)}}{\mu-\mu^{-1}}.
\end{gathered}
\label{eq:S43}
\end{equation}

\begin{figure}[!tb]
\centering
\includegraphics[width=\linewidth]{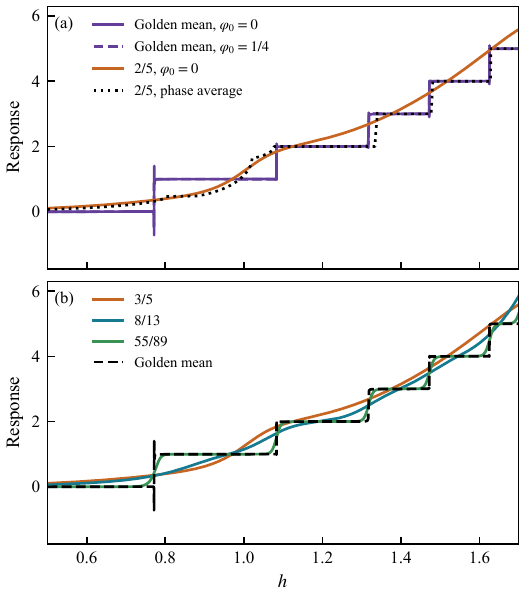}
\caption{Comparison of periodic and quasiperiodic sequences,
$\omega_0\tau=9.03$. (a)~Dependence on the initial phase. Quasiperiodic
curves: forward-input response $s_N^{(+)}$ of the golden-mean sequence,
$N=8192$, $\varphi_0=0$ (solid) and $1/4$ (dashed). Periodic curves:
Floquet response $s_{\rm per}$ of the cell $2/5$ at $\varphi_0=0$ (solid)
and its average over $\varphi_0$ (dotted). (b)~Approach as the cell
becomes larger. Floquet responses $s_{\rm per}$ of the cells $3/5$,
$8/13$, and $55/89$ at $\varphi_0=0$; each curve is the limit of infinite
repetition of the $q$-layer cell. Dashed: golden-mean sequence with
$N=8192$ and $\varphi_0=0$.}
\label{fig:S5}
\end{figure}

A periodic sequence passes only through the $q$ phases on its own orbit
and does not sweep all phases, so, however long it is, its response
depends on the initial phase. For example, for the cell $2/5$ at
$h=1.20$, the response varies with $\varphi_0$ between $0.973$ and $2.348$
[Fig.~\ref{fig:S5}(a)]. An integer can appear at most after an average
over the initial phase. If $|\mu|>1$ for every initial phase, $\mu$ is an
analytic function of the initial phase and, like the trace, has the period
$1/q$, so its winding number over the whole phase circle is an integer
multiple of $q$, and therefore
\begin{equation}
\int_0^1 s_{\rm per}\,\dd\varphi_0=-\frac1q\,\wind\mu\in\mathbb Z.
\label{eq:S44}
\end{equation}

For the cell $2/5$ at $h=1.20$, this average equals 2. But ``$|\mu|>1$ for
every initial phase'' does not always hold. As soon as some initial phases
fall into a pass band, where $|\mu|=1$ and the wave neither grows nor
decays over the cell, the averaged slope is no longer an integer and
changes continuously with $h$. For example, this is the case for the cell
$2/5$ over the whole range $h<1.08$; at $h=0.90$ the averaged slope is
$0.525$, whereas the quasiperiodic sequence gives 1 there. These two
problems of periodic sequences, a response that depends on the initial
phase and an average that may still not be an integer, do not exist for
an irrational rotation number. Avila's theorem guarantees that the
acceleration is an integer at every $h$ (Sec.~S2\,A), so no interval of
non-integer values appears; Sec.~S1\,B has explained that the phases
passed by a single quasiperiodic sequence sweep the whole phase circle
uniformly, so a single sequence performs the average over the initial
phase by itself, and its growth rate does not depend on the initial phase.
In Fig.~\ref{fig:S5}(a), the two quasiperiodic curves for $\varphi_0=0$
and $1/4$ give the same plateaus. Periodic sequences can only approach
this behavior step by step: as the cell becomes larger, the Floquet
response at a fixed initial phase develops flatter and flatter intervals
that come closer and closer to the plateaus of the quasiperiodic sequence,
but it does not form exact integer plateaus [Fig.~\ref{fig:S5}(b)].

\begin{figure}[!tb]
\centering
\includegraphics[width=0.94\linewidth]{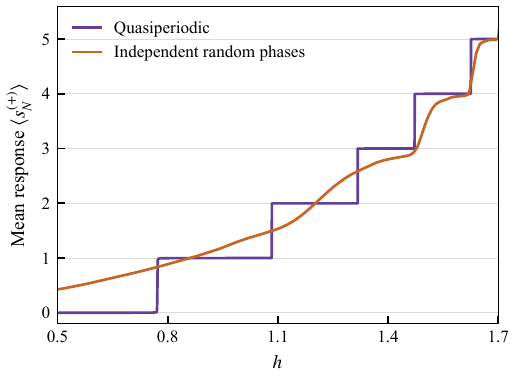}
\caption{Responses of quasiperiodic and random orders along the same $h$
path, $N=8192$, $\omega_0\tau=9.03$, forward input. Angle brackets denote
averages of the forward-input response $s_N^{(+)}$: over 512 initial
phases for the quasiperiodic sequence and over 1024 independent sequences
for the random sequence; each response is first computed from the
logarithm of the output norm and then averaged. The shading is one
standard error of the random average.}
\label{fig:S6}
\end{figure}

To examine the role of the order of the layers by itself, we replace
$\varphi_0+m\alpha$ by independent random phases $\xi_m$ uniformly
distributed on $[0,1)$:
\begin{equation}
\varepsilon_m=\varepsilon_R+\delta\varepsilon\sin(2\pi\xi_m+\ii h).
\label{eq:S45}
\end{equation}

Each $\xi_m$ has the same distribution as the layer phases of the
quasiperiodic sequence, and the permittivity depends on $h$ through the
same function; only the correlation between the phases of different layers
is removed. Each random sequence is kept fixed while $h$ is scanned. The
wave still grows exponentially, but the response changes continuously
with $h$, and the plateaus disappear (Fig.~\ref{fig:S6}). Earlier work
studied disordered photonic time crystals with randomly distributed real
permittivity~\cite{Sharabi2021}, where the strength of the disorder is
tuned; here the random sequence changes along the same $h$ path as the
quasiperiodic sequence. Of the three orders compared here, only the
quasiperiodic order gives integer plateaus for a single sequence (fixed
initial phase): the random order removes the plateaus, and the periodic
order keeps an integer at most after an average over the initial phase.
Therefore the integer plateaus remain unchanged under perturbations that
preserve analyticity and the irrational rotation number (Sec.~S4\,A), but
they cannot exist without the quasiperiodic order.

\section{Output of finite sequences}

\subsection{Output ports and finite-length corrections}

\begin{figure}[!tb]
\centering
\includegraphics[width=0.94\linewidth]{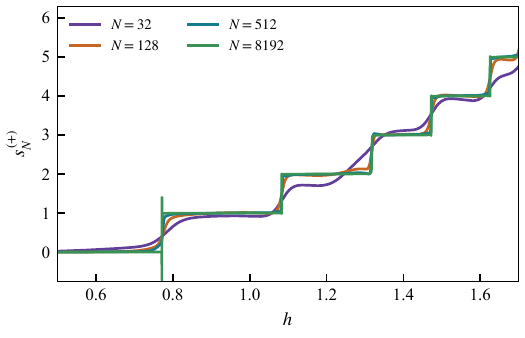}
\caption{Finite-length corrections, $\omega_0\tau=9.03$, $\varphi_0=0$.
Forward-input response $s_N^{(+)}$ of the golden-mean sequence at four
lengths.}
\label{fig:S7}
\end{figure}

A real modulation sequence is always finite: before the modulation starts
and after it ends, the medium is the reference medium, the initial phase
is fixed at $\varphi$, and the input is a given state $v_{\rm in}$. Its
logarithmic growth rate per layer, averaged over the $N$ layers, is
\begin{equation}
g_N(\varphi,h;v_{\rm in})=\frac1N\ln\frac{\|M_N(\varphi,h)v_{\rm in}\|}{\|v_{\rm in}\|},
\label{eq:S46}
\end{equation}
where the norm is the Euclidean norm of a vector. After the modulation
ends, the medium returns to the reference medium, $z=1$. Taking any
reference time $s_0$ in this medium, the displacement amplitude after it
is a superposition of two waves traveling along $+x$ and along $-x$:
\begin{equation}
\begin{gathered}
D(s)=A_+\ee^{-\ii(s-s_0)}+A_-\ee^{\ii(s-s_0)},\\
A_\pm=\tfrac12\bigl[D(s_0)\pm\ii D'(s_0)\bigr].
\end{gathered}
\label{eq:S47}
\end{equation}
We call these two waves the forward and backward waves; they travel along
$+x$ and $-x$, respectively, and correspond to two output ports. The
forward-wave input of unit amplitude is $v_{\rm in}=(1,1)^{\mathsf T}$,
and we write its growth rate as $g_N^{(+)}$. At the end of the modulation,
the state $v_N=(D_N,\ii D'_N)^{\mathsf T}$ has the following amplitudes and
powers at the two ports:
\begin{equation}
\begin{gathered}
t_N=\frac{D_N+\ii D'_N}{2},\qquad r_N=\frac{D_N-\ii D'_N}{2},\\
\mathcal T_N=|t_N|^2,\quad\mathcal R_N=|r_N|^2.
\end{gathered}
\label{eq:S48}
\end{equation}
The two ports are normalized in the same medium, so the total gain is
\begin{equation}
\begin{gathered}
G_N\equiv\mathcal T_N+\mathcal R_N=\frac{|D_N|^2+|D'_N|^2}{2}\\
=\frac{\|M_Nv_{\rm in}\|^2}{\|v_{\rm in}\|^2},\\
g_N^{(+)}=\frac{\ln G_N}{2N}.
\end{gathered}
\label{eq:S49}
\end{equation}
That is, the growth rate of a forward input is set by the sum of the
output powers of the two temporal ports. Its derivative with respect to
$h$ is the forward-input response used in Sec.~S4\,A:
\begin{equation}
s_N^{(+)}(\varphi,h)=\partial_hg_N^{(+)}(\varphi,h).
\label{eq:S50}
\end{equation}
Figures~\ref{fig:S4}, \ref{fig:S5}, and~\ref{fig:S7} all show this
quantity.

The difference between the powers of the two ports obeys a balance
relation, which shows a feature that complex modulation brings and that
Hermitian time modulation does not have. We define the flux
$\Phi(s)=-\operatorname{Im}(D^*D')$. In the reference medium,
$\Phi=|A_+|^2-|A_-|^2$, the forward power minus the backward power; by
Eq.~\eqref{eq:S3}, $\dd\Phi/\dd s=\operatorname{Im}(z^2)\,|D|^2$. The flux $\Phi$ is
continuous at the interfaces and equals 1 for a forward-wave input, so
\begin{equation}
\mathcal T_N-\mathcal R_N=1+\int_0^{N\omega_0\tau}
\operatorname{Im}\!\left[\frac{\varepsilon_R}{\varepsilon(s)}\right]|D(s)|^2\,\dd s.
\label{eq:S51}
\end{equation}
For a real permittivity the integral vanishes and
$\mathcal T_N-\mathcal R_N=1$: in Hermitian photonic time crystals and time
quasicrystals, however the layers are ordered, the difference between the
forward and backward powers always equals the input power, and the two
ports can only grow together. This also shows that $\det S_m=1$ does not
imply the conservation of electromagnetic energy: even for a real
permittivity, only $\mathcal T_N-\mathcal R_N$ is conserved, while the
total power $\mathcal T_N+\mathcal R_N$ can still grow exponentially, and
the energy for the growth is supplied by the modulation. For a complex
permittivity, the integrand is a product of two factors: one is
$\operatorname{Im}(\varepsilon_R/\varepsilon)
=-\varepsilon_R\operatorname{Im}\varepsilon/|\varepsilon|^2$, which gives
the gain or loss of each layer; the other is the field intensity $|D|^2$.
The phase average of the first factor is zero, but the field intensity is
not the same in the gain layers and in the loss layers, so the integral is
in general nonzero, and $\mathcal T_N-\mathcal R_N$ deviates from 1. Thus
$\mathcal T_N-\mathcal R_N-1$ equals the sum, over the whole sequence, of
the gain and loss of each layer weighted by the field intensity, and it is
nonzero only under complex modulation.

How far is the response of a finite sequence from the integer plateau?
This paragraph shows that the error decays as $1/N$. Let $h$ lie in an
integer phase with $g_\infty>0$, and let the input be a forward wave that
has a nonzero component in the growing direction at the initial phase. In
each layer this growing component is multiplied by the multiplier
$\lambda$ [Eq.~\eqref{eq:S12}], and the decaying component becomes smaller
and smaller relative to it. The growth rate of a finite sequence therefore
equals the average of $\ln|\lambda|$ over the layers plus a bounded
quantity divided by $N$. We write
$F_h(\varphi)=\ln|\lambda(\varphi+\ii h/2\pi)|-g_\infty(h)$, the part of
$\ln|\lambda|$ of one layer that deviates from the bulk growth rate; its
phase average is zero [Eq.~\eqref{eq:S14}]. Then
\begin{equation}
\begin{gathered}
g_N^{(+)}(\varphi,h)-g_\infty(h)\\ =\frac1N\sum_{m=0}^{N-1}F_h(\varphi+m\alpha)
+\frac{\beta_N(\varphi,h)}{N},
\end{gathered}
\label{eq:S52}
\end{equation}
where the bounded term $\beta_N$ comes from the projection of the input
onto the growing direction, from the end-point term in
Eq.~\eqref{eq:S13}, and from the decaying component. The size of the error
therefore depends on how the sum $\sum_mF_h(\varphi+m\alpha)$ grows with
$N$. The function $F_h$ has zero average, so some terms are positive and some are
negative. If the layer phases were independent random numbers, this sum
would in general grow as $\sqrt N$, and the error would decay only as
$1/\sqrt N$. For the golden mean, this sum is bounded, so the error decays
as $1/N$.

That this sum is bounded in $N$ for the golden mean is a classical result
for quasiperiodic rotations~\cite{KatokHasselblatt1995}: for rotation
numbers that are hard to approximate by rational numbers, such as the
golden mean, an analytic function with zero average can be written as a
difference of a periodic function $\chi_h$,
$F_h(\varphi)=\chi_h(\varphi+\alpha)-\chi_h(\varphi)$, and in the sum along
the orbit adjacent terms cancel, leaving only the bounded
$\chi_h(\varphi+N\alpha)-\chi_h(\varphi)$. Random phases have no such
cancellation. The same conclusion holds after differentiation with
respect to $h$. Therefore, as long as $h$ stays in an integer phase and the
input always has a growing component, the differences between the growth
rate and the response of a finite sequence and their bulk values both
decay as $1/N$.

Figure~\ref{fig:S7} fixes the initial phase and the input and increases
the length $N$ of the sequence: within each integer phase the response
becomes flatter and flatter, and the transitions between phases become
steeper and steeper, consistent with the $1/N$ convergence above.

\subsection{Response to a two-frequency input}

Figure~4 of the main text sends two frequencies into the same modulation
sequence at the same time. This section is concerned with how the ratio
$P_2/P_1$ of the output powers at the two frequencies changes with $h$,
and shows that it is set by the difference between the integers of the
two frequencies. At a given $h$, the two frequencies may lie in different
integer phases, with the bulk integers $n_1$ and $n_2$. By Sec.~S5\,A, when
$h$ changes by $\Delta h$, the output power at frequency $j$ changes by a
factor of about $\ee^{2Nn_j\Delta h}$, and the correction due to the
finite length does not grow with $N$, so
\begin{equation}
\Delta\ln\frac{P_2}{P_1}\simeq2N(n_2-n_1)\,\Delta h.
\label{eq:S53}
\end{equation}
If the two frequencies lie in the same integer phase, changing $h$
amplifies both by the same factor, and the power ratio does not change. If
they lie in different integer phases, the power ratio changes by a factor
$\ee^{2N(n_2-n_1)\Delta h}$ for every change $\Delta h$ of $h$: the larger
the integer difference and the longer the sequence, the larger the change
of the power ratio caused by the same $\Delta h$.

\begin{figure}[!tb]
\centering
\includegraphics[width=\linewidth]{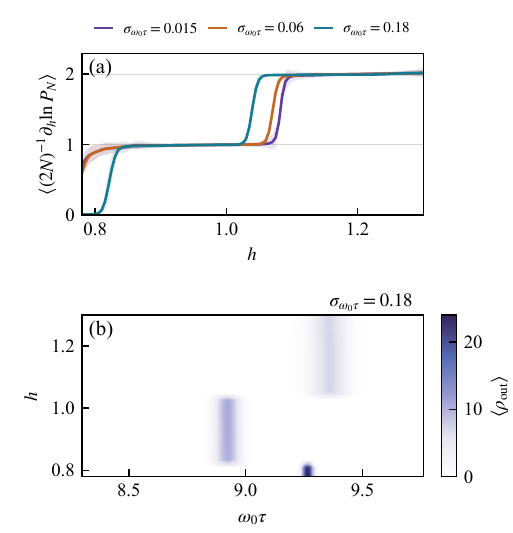}
\caption{Band-power response of the golden-mean sequence for inputs of
finite bandwidth, band center $\omega_0\tau=9.03$, $N=144$, averaged over
32 equally spaced initial phases. (a)~Normalized response
$(2N)^{-1}\partial_h\ln P$ of the band power $P$ (sum of the two ports)
for three bandwidths $\sigma_{\omega_0\tau}$; the shading is the standard
deviation over the initial phase, and horizontal lines mark the integers.
(b)~Normalized distribution of the output power within the band versus $h$
and $\omega_0\tau$ for $\sigma_{\omega_0\tau}=0.18$; at each $h$ the
distribution integrates to 1, and all values of $h$ share the same color
scale.}
\label{fig:S8}
\end{figure}

When the two frequencies are input coherently, the total intensity at the
output oscillates at their difference frequency; this is the beat. The
strength of the beat is measured by the interference visibility, that is,
$(I_{\max}-I_{\min})/(I_{\max}+I_{\min})$ of the intensity oscillation.
The visibility is at most $2\sqrt{P_1P_2}/(P_1+P_2)$, which equals 1 when
the two frequencies have equal powers. If one frequency is amplified by a
factor $\ee^{2N(n_2-n_1)\Delta h}$ more than the other, this upper bound
decreases exponentially with $N$, the beat disappears, and the output
effectively contains only one frequency. The field maps in
Figs.~4(c)--4(e) of the main text show exactly this.

A real input has a finite bandwidth; $P_j$ is then the total output power
of band $j$. The gain differs between the frequencies within a band, so
the rate of change of the band-power ratio equals the difference between
the two bands of the averaged response, where in each band the responses
$s_N^{(+)}$ of its frequencies are averaged with their output powers as
weights. If each band lies
entirely within one integer phase, this difference tends to $n_2-n_1$ for
large $N$; if a band crosses a phase boundary, some frequencies in the
band take one integer and the others take the adjacent integer, the result
lies between the two, and the phase boundary is therefore broadened by the
bandwidth. Figure~\ref{fig:S8} uses a single band as an example: the
larger the bandwidth, the more gradual the transition of the band-power
response near a phase boundary, and the position of the transition also
moves with the weights of the output spectrum.

Equation~\eqref{eq:S53} can also be used to estimate the result of Fig.~4
of the main text. Between the two settings compared in the main text, the
second band crosses a phase boundary into $n=2$, while the first band
stays at $n=1$. Summing $n_2-n_1$ over this range of $h$,
Eq.~\eqref{eq:S53} predicts an increase of $7.27$ in the logarithm of the
power ratio; the propagation of the full input spectrum gives $6.90$,
which is the $30$~dB of the main text. The two differ by $5\%$, which
comes from the corrections due to the finite length and the finite
bandwidth.

\begin{figure}[!tb]
\centering
\includegraphics[width=\linewidth]{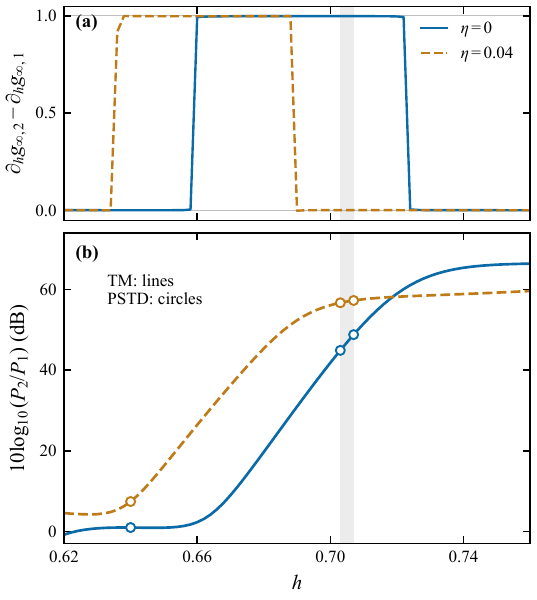}
\caption{Two-frequency response under waveform distortion, for the two
bands of Fig.~4 of the main text, $\eta=0$ (solid) and $0.04$ (dashed).
(a)~Difference between the bulk responses at the two band centers.
(b)~Output power ratio $10\log_{10}(P_2/P_1)$ for $N=120$ and
$\varphi_0=0$: lines are transfer-matrix calculations with the full input
spectrum (TM), and circles are pseudospectral time-domain simulations
(PSTD), which solve Maxwell's equations of the same medium directly in
space and time (Sec.~S6). The gray band marks the window
$h\in[0.703,0.707]$.}
\label{fig:S9}
\end{figure}

A distortion of the modulation waveform moves the phase boundaries
(Sec.~S4\,A) and thereby moves the range of $h$ in which ``the two bands
take different integers''. We take the second harmonic of Sec.~S4\,A (with
the phase difference $\psi=0.37$ used in Fig.~3(e) of the main text) and
increase its amplitude $\eta$ from 0 to $0.04$; for $|h|\le1.1$ the
modulus of the permittivity stays above $1.54$, and the modulation stays
inside the analytic strip. For $\eta=0$, the centers of the second and
first bands enter $n=2$ at $h=0.6597$ and $0.7227$, respectively; for
$\eta=0.04$, these two boundaries move to $0.6359$ and $0.6895$. The
interval with $n_2-n_1=1$ therefore moves to smaller $h$, and its width
shrinks from $0.063$ to $0.054$ [Fig.~\ref{fig:S9}(a)].

This shift can open or close the relative amplification within a given
window of $h$. In the window $h\in[0.703,0.707]$, the two bands lie in
$n=1$ and $n=2$, respectively, for $\eta=0$, and both lie in $n=2$ for
$\eta=0.04$. We obtain the integer difference from the difference of the
logarithm of the power ratio between the two ends of the window divided by
$2N\Delta h$, that is, from the average rate of change of the power ratio;
its bulk values are 1 and 0, respectively. A sequence with $N=120$ gives
$0.936$ and $0.138$, and $N=2048$ gives $1.007$ and $0.006$, both
approaching the bulk values [Fig.~\ref{fig:S9}(b) and Sec.~S6]. Therefore
a second harmonic with an amplitude of only $4\%$ pushes the first band
across its phase boundary and closes the relative amplification in this
window. Conversely, the modulation waveform itself is also a means of
control: changing the amplitude or the phase of the harmonic moves the
phase boundaries and thus selects the windows of $h$ in which the two
frequencies are amplified differently.

\section{Numerical methods and scope of the model}

This section explains how the numerical results in the figures are
obtained and under what conditions these results apply to real media.
Section~S6\,A describes the transfer-matrix calculations, Sec.~S6\,B
describes a time-domain simulation that is independent of them, and
Sec.~S6\,C discusses the scope of the constitutive model. The physical
parameters of each figure are given in its caption.

\subsection{Transfer-matrix calculations}

\paragraph{Growth rates and responses.}
Except for the time-domain simulation, all calculations are based on
ordered products of layer matrices [Eq.~\eqref{eq:S4}]. To avoid overflow,
the product is repeatedly divided by positive factors during the
calculation, and the logarithms of these factors are accumulated. The
bulk growth rate is the logarithm of the largest singular value of the
product divided by $N$, averaged over $N_\varphi$ equally spaced initial
phases; the growth rate for a given input is obtained from the norm of the
propagated state. The response, that is, the derivative of the growth rate
with respect to $h$, is not computed by numerical differences; instead,
the tangent vector $\widetilde v_m=\partial_hv_m$ is propagated together
with the state:
\begin{equation}
\begin{gathered}
\widetilde v_{m+1}=S_m\widetilde v_m+(\partial_hS_m)v_m,\\
\partial_hg_N=\frac{\operatorname{Re}\bigl(v_N^\dagger\widetilde v_N\bigr)}{N\|v_N\|^2}.
\end{gathered}
\label{eq:S54}
\end{equation}
For an input that does not depend on $h$, $\widetilde v_0=0$. The
rescaling does not affect the result: $v_m$ and $\widetilde v_m$ are
divided by the same positive factor each time and still satisfy the same
linear recursion, and the right side of Eq.~\eqref{eq:S54} does not change
when the numerator and the denominator are multiplied by the same positive
number. The results of the tangent vector were checked against central
differences. All responses in the figures are the computed values
themselves, without smoothing or rounding. The edge value $g_\star$ of
Sec.~S3\,B is also computed with such products: each layer matrix is first
divided by the modulus of its larger eigenvalue,
$\exp(\omega_0\tau|\operatorname{Im}z|)$, and then propagated; the
resulting growth rate is the remainder $R$, and adding $J$ obtained by
integrating Eq.~\eqref{eq:S33} gives $g_\star$.

\paragraph{Integers and phase boundaries.}
Finite $N$ and $N_\varphi$ can only give responses close to integers, so a
criterion is needed to decide which points count as integers. We count a
point as an integer when it satisfies two conditions at the same time: the
response is within $0.02$ of the nearest integer, and halving $N$ or
$N_\varphi$ changes the response by less than $0.01$, that is, the result
has converged. The colored segments in Fig.~2(b) and the coloring in
Fig.~2(c) of the main text show only these points. Between two integer
phases the response changes continuously from $n$ to $n+1$, and we take
the position where it crosses $n+\tfrac12$ as the phase boundary. The
markers in Fig.~3(d) of the main text are obtained in this way, with
$N=4096$ for the transfer-matrix calculation and $N=512$ for the
time-domain simulation. The phase boundaries predicted from the Fourier
coefficients do not use this criterion; their convergence with $q$ is given
in Sec.~S3\,A.

\paragraph{Two-frequency input.}
The input of Fig.~4 of the main text and of Sec.~S5\,B is not a single
frequency but two bands of finite width, centered at
$\omega_{0,1}\tau=15.54$ and $\omega_{0,2}\tau=15.82$. The two bands have
equal input powers, the same for all $h$. On the discrete frequency grid,
the incident power of each mode is
\begin{equation}
\begin{gathered}
w_j(\omega_0\tau)\propto\exp\!\left[-\nu^2-\frac{1}{1-\nu^2/16}\right]\ (|\nu|<4),\\ \nu=\frac{\omega_0\tau-\omega_{0,j}\tau}{0.005},
\end{gathered}
\label{eq:S55}
\end{equation}
and zero otherwise. Such an input contains 52 modes in total, and the
complex amplitudes of all modes have the same phase. The output power of
each mode is $\mathcal T_N+\mathcal R_N$ times its input power
[Eq.~\eqref{eq:S49}], and $P_j$ is the sum of the output powers of the
modes in band $j$.

\subsection{Pseudospectral time-domain simulation}

\paragraph{Method.}
To test the results independently of the transfer matrix, we use the
pseudospectral time-domain (PSTD) method, which solves Maxwell's equations
directly in space and time and computes spatial derivatives with Fourier
transforms~\cite{Liu1997PSTD,Liu1999PSTD}. With the dimensionless
coordinate $\bar x=x/(v_0\tau)$, the time $T=t/\tau$, and the scaled
magnetic field $\mathcal B=v_0\varepsilon_0\varepsilon_RB_z$, where
$v_0=c/\sqrt{\varepsilon_R}$, Eq.~\eqref{eq:S1} becomes, for the full
fields $D(\bar x,T)$ and $\mathcal B(\bar x,T)$ (rather than single Fourier
amplitudes),
\begin{equation}
\partial_TD=-\partial_{\bar x}\mathcal B,\qquad
\partial_T\mathcal B=-\frac{\varepsilon_R}{\varepsilon(T)}\,\partial_{\bar x}D.
\label{eq:S56}
\end{equation}
A single Fourier component $\ee^{\ii\omega_0\tau\bar x}$ satisfies
$\mathcal B=\ii D'$ and evolves with the layer matrices of
Eq.~\eqref{eq:S4}, so the two methods describe the same physical problem.
We write the fields as a carrier times an envelope,
$D=\ee^{\ii K_c\bar x}\widetilde D$ and
$\mathcal B=\ee^{\ii K_c\bar x}\widetilde{\mathcal B}$, where the carrier
wavenumber $K_c$ is taken as a value of $\omega_0\tau$ near the center of
the input spectrum. On a periodic grid of length $L$ with $N_x$ nodes, the
spatial derivative is computed through the envelope,
$\partial_{\bar x}(\ee^{\ii K_c\bar x}\widetilde D)
=\ee^{\ii K_c\bar x}\,\mathcal D\widetilde D$, where
\begin{equation}
\begin{gathered}
\mathcal D=\mathcal F^{-1}\,\ii(K_c+\bar k_\ell)\,\mathcal F,\\
\bar k_\ell=\frac{2\pi\ell}{L},\\
\ell=-\frac{N_x}{2},\ldots,\frac{N_x}{2}-1,
\end{gathered}
\label{eq:S57}
\end{equation}
and $\mathcal F$ is the discrete Fourier transform. The time integration
uses the classical fourth-order Runge--Kutta scheme, and every switching of
the permittivity falls on a boundary between time steps, where $D$ and
$\mathcal B$ are continuous. All modes on the grid take part in the
evolution, and no filtering or projection is applied during the
propagation. The medium is spatially uniform, so the spatial derivative is
exact for every mode on the grid, and the discretization error comes only
from the time integration.

\paragraph{Accuracy.}
The simulation of Fig.~4 of the main text uses $K_c=15.68$ and $L=4096$;
the field maps use $N_x=4096$ and 512 time steps per layer, and the scan
of the power ratio uses $N_x=2048$ and 256 steps per layer. Over the whole
scan, the power ratio from the time-domain simulation agrees with the
transfer-matrix calculation for the same input spectrum to within
$0.002$~dB. When the number of time steps is doubled, the error of the
complex fields shrinks to about $1/16$, as expected for a fourth-order
scheme; the power in the modes not excited by the input always stays below
$10^{-20}$ of the total power.

\subsection{Scope of the constitutive model}

\paragraph{What the model assumes.}
All calculations in this work are based on an idealized constitutive
relation: the permittivity switches instantaneously between layers and is
the same for all frequencies, as in earlier studies of non-Hermitian
time-varying media~\cite{Wang2018}. This assumption has a direct
consequence. Within a temporal layer, a wave of wavenumber $k$ oscillates
at the frequency $c|k|/\sqrt{\varepsilon}$; when $\varepsilon$ is complex,
this frequency is also complex, and its imaginary part is the gain or
decay per unit time, which is proportional to $|k|$. In this model,
therefore, the larger the wavenumber, the larger the gain within each
layer, without an upper limit. Real media cannot do this; their gain
exists only over a finite range of wavenumbers.

\paragraph{What real media need to satisfy.}
This difference does not prevent the observation of the integer response
in real media. The medium is spatially uniform, so different wavenumbers
do not couple and evolve independently. A real medium therefore only
needs to approximate this model over the range of wavenumbers occupied by
the input: the real and imaginary parts of the permittivity oscillate with
the modulation phase as in Eq.~\eqref{eq:S6}, with amplitudes set by the
same $h$. The calculations for finite sequences in Sec.~S5 and in Fig.~4
of the main text use exactly such inputs of finite bandwidth and finite
numbers of layers. In addition, the imaginary part of the modulation must
be large enough. For the parameters of this work, the integer phase $n=1$
at $\omega_0\tau=9.03$ begins at $h\approx0.77$, where the amplitude of
the imaginary part is $\delta\varepsilon\sinh h\approx0.85$, about one
quarter of $\varepsilon_R$; over the frequency range of Fig.~2(c) of the
main text, the amplitude of the imaginary part needed to enter $n=1$ is
always above $0.7$. Causality requires the gain to come with dispersion,
and gain media also have noise and saturation; the model of this work does
not include these effects. Section~S7 gives a circuit that meets the above
requirements.

\section{Circuit realization}

This section gives a circuit that realizes the model of this work exactly
with ideal components; all the components it needs are common circuit
components.

\paragraph{Role of the imaginary part of the permittivity.}
The most direct approach is to provide loss with resistors and gain with
amplifiers (negative resistance) in each layer, but this does not give the
model of this work. A shunt conductance or negative conductance gives the
same gain or loss to the forward wave and to the backward wave produced by
time reflection, so the determinant of the layer matrix is not 1. The
layer matrix of this work instead belongs to $\mathrm{SL}(2,\mathbb C)$
(Sec.~S1\,A): within a layer, when the forward wave grows at some rate, the
backward wave decays at the same rate. Therefore the imaginary part of the
permittivity is not an ordinary gain or loss; it acts oppositely on the
forward and backward waves. The circuit below realizes this with an
antisymmetric coupling between two rings formed by lumped transmission
lines.

\begin{figure}[!tb]
\centering
\includegraphics[width=.80\linewidth]{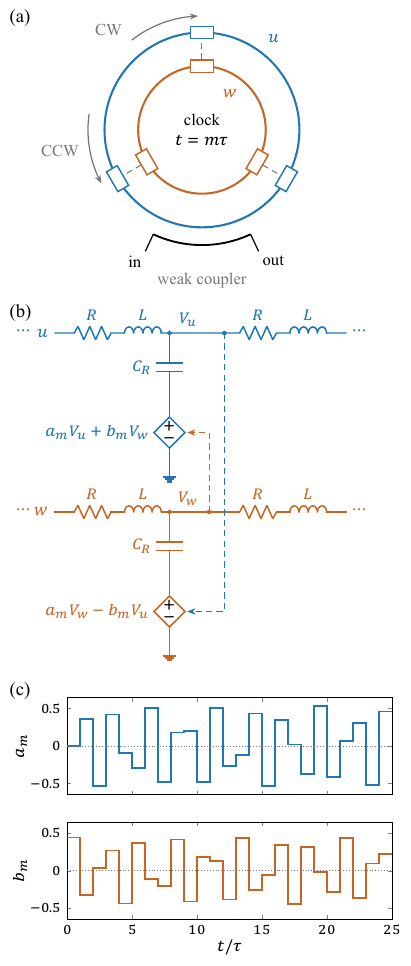}
\caption{Circuit that realizes the model of this work. (a)~Two identical
lumped transmission-line rings $u$ (outer) and $w$ (inner), each
consisting of $M$ cells ($M=3$ in the figure); the dashed lines indicate
the coupling between the cells at the same position in the two rings. All
cells switch at the same times $t=m\tau$, controlled by one clock. A
weakly coupled directional coupler on ring $u$ injects the input and reads
out separately the waves traveling clockwise (CW) and counterclockwise
(CCW). (b)~One cell: a resistor $R$ in series with an inductor $L$, and a
shunt branch to ground consisting of a fixed capacitor $C_R$ in series
with a controlled voltage source. The controlled sources sense the node
voltages to ground; the source voltage is $a_mV_u+b_mV_w$ in ring $u$ and
$a_mV_w-b_mV_u$ in ring $w$, and the dashed arrows indicate the voltages
taken from the other ring. (c)~Gains $a_m$ and $b_m$ of the first 25
layers at $h=1.20$ and $\varphi_0=0$, given by Eq.~\eqref{eq:S59}; here
$|a_m|\le0.53$ and $|b_m|\le0.44$.}
\label{fig:S10}
\end{figure}

\paragraph{Circuit.}
The circuit consists of two identical lumped transmission-line rings, $u$
and $w$. Each ring is formed by $M$ identical cells connected end to end
and therefore has a discrete translation symmetry [Fig.~\ref{fig:S10}(a)].
Each cell has a resistor $R$ and an inductor $L$ in series, and a shunt
branch to ground; the branch consists of a fixed capacitor $C_R$ in series
with a controlled voltage source [Fig.~\ref{fig:S10}(b)]. The controlled
sources sense the node voltages to ground, $V_u$ and $V_w$, at the same
position in the two rings: the voltage of the controlled source in ring
$u$ is $a_mV_u+b_mV_w$, and that in ring $w$ is $a_mV_w-b_mV_u$. Each
controlled source can be realized with two analog multipliers. All cells
are controlled by the same clock: in layer $m$, the controlled sources of
all cells use the same pair of gains $a_m$ and $b_m$
[Fig.~\ref{fig:S10}(c)], so the whole circuit needs only two control
waveforms. No component value in the circuit changes with time; the time
modulation is carried out entirely by the gains of the controlled sources.
A temporal layer is a time interval set by the clock, so the number of
layers depends only on how long the modulation lasts and not on the size
of the rings.

\paragraph{Why the circuit realizes the model.}
We write the charges on the capacitors at the same position in the two
rings as $q_u$ and $q_w$, and the node voltages as $V_u$ and $V_w$, and we
combine them into the complex numbers $D=q_u+\ii q_w$ and
$V=V_u+\ii V_w$. The voltages of the two controlled sources combine to
$(a_m-\ii b_m)V$, and the voltage across a capacitor equals the node
voltage minus the voltage of the controlled source, so
$D=C_R(1-a_m+\ii b_m)V$. That is, for the complex combination, the shunt
branch acts as a complex capacitance $C_R(1-a_m+\ii b_m)$. The inductances
and resistances of the two rings are the same, so the evolution of $D$ on
the rings is the same as on a single ring whose capacitance is this
complex capacitance. The modes of the ring differ by a phase factor
$\ee^{\ii k}$ between adjacent cells, with $k=2\pi\times\text{integer}/M$,
and each mode evolves independently. By Kirchhoff's laws, each mode
satisfies
\begin{equation}
\ddot D+\frac RL\,\dot D=-\frac{\omega_0^2}{1-a_m+\ii b_m}\,D,\qquad
\omega_0=\frac{2}{\sqrt{LC_R}}\sin\frac k2.
\label{eq:S58}
\end{equation}
Consider first $R=0$. Written with $\omega_0$, Eq.~\eqref{eq:S2} reads
$\ddot D_k=-\omega_0^2(\varepsilon_R/\varepsilon_m)D_k$, and the two
equations agree when the complex capacitance is proportional to the
permittivity, $1-a_m+\ii b_m=\varepsilon_m/\varepsilon_R$. By
Eq.~\eqref{eq:S6},
\begin{equation}
a_m=-\frac{\delta\varepsilon}{\varepsilon_R}\cosh h\,\sin X_m,\qquad
b_m=\frac{\delta\varepsilon}{\varepsilon_R}\sinh h\,\cos X_m.
\label{eq:S59}
\end{equation}
The two gains are exactly the real and imaginary parts of the permittivity
modulation, each divided by $\varepsilon_R$: $a_m$ realizes the modulation
of the real part, and $b_m$ that of the imaginary part. The imaginary part
of the permittivity thus becomes an antisymmetric coupling between the two
rings, and the energy it needs is supplied by the controlled sources. Both
$a_m$ and $b_m$ are real and change sign from layer to layer; a sign
change of $b_m$ corresponds to the imaginary part of the permittivity
alternating between gain and loss, and in the circuit it only means that
the gain of the controlled source takes a negative value, which a
four-quadrant multiplier realizes directly. The amplitudes of the two
gains are set by the same $h$, and their phases differ by a quarter
period, the same relation as between the real and imaginary parts in
Eq.~\eqref{eq:S6}. At a switching time, the charges on the capacitors and
the currents in the inductors are continuous, and a jump of the source
voltage only changes the node voltage, so $D$ and $\dot D$ are continuous,
the same matching conditions as in Sec.~S1\,A. Therefore the layer matrix
of each ring mode in the circuit is Eq.~\eqref{eq:S4}.

\begin{figure}[!tb]
\centering
\includegraphics[width=\linewidth]{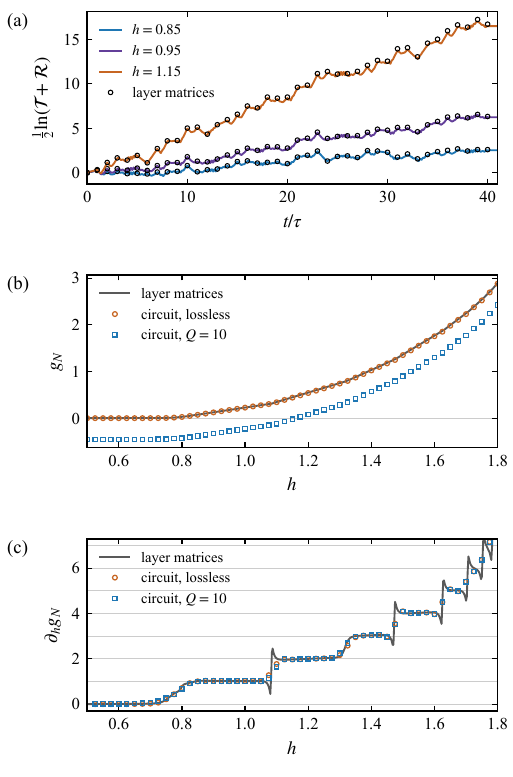}
\caption{Transient simulation of the circuit of Fig.~\ref{fig:S10}
($M=3$) compared with the layer matrices [Eq.~\eqref{eq:S4}].
(a)~$\tfrac12\ln(\mathcal T+\mathcal R)$ versus time without loss, for
$\varphi_0=0.3$, a forward input, and three values of $h$; the curves are
the circuit simulation, and the circles are the results of the layer
matrices at the end of each layer. (b)~Growth rate $g_N$ for $N=60$,
averaged over 8 initial phases: without loss (circles) and with $Q=10$
(squares). (c)~Response $\partial_hg_N$ obtained from (b); gray lines mark
the integers. Loss lowers the growth rate as a whole but does not change
the integer plateaus. At $h=0.90$, $1.20$, and $1.40$ ($N=60$, 16 initial
phases), the response differs from that of the ideal circuit by at most
$0.06$ when the input is injected into ring $u$ only, when each gain
switching takes $5\%$ of a layer, when the bandwidth of the controlled
sources is 10 times the operating frequency, when all gains are $2\%$ too
large, or when the loss is $Q=10$; with random deviations of $1\%$ in the
inductors and capacitors of the cells, the difference is about $0.05$.}
\label{fig:S11}
\end{figure}

\paragraph{Loss does not change the integer.}
The loss from the series resistors does not affect the integer response.
Writing $D=\ee^{-Rt/2L}\widetilde D$, $\widetilde D$ satisfies the equation
for $R=0$, except that $\varepsilon_R/\varepsilon_m$ in it is reduced as a
whole by the constant $1/(4Q^2)$, where $Q=\omega_0L/R$ is the quality
factor of the inductor at the operating frequency. Loss therefore has two
effects: first, the growth rate of every layer is reduced by the same
constant $\omega_0\tau/(2Q)$, which does not depend on $h$; second, the
layer matrix changes slightly, by an amount of order $1/(4Q^2)$. The first
effect cancels in the derivative with respect to $h$; the effect of the
second on the response is negligible [Fig.~\ref{fig:S11}(c)]. This is
useful for experiments: a suitable choice of $R$ lowers the net gain
without changing the slope to be measured.

\paragraph{Circuit simulation.}
To test this circuit directly, we performed a transient simulation of it.
The simulation writes the circuit equations element by element (modified
nodal analysis) and integrates them with the trapezoidal rule, with 800
time steps per layer, and each switching of the source gains is taken as a
linear transition that lasts $0.5\%$ of $\tau$; the simulation uses
neither the complex combination $D$ nor the layer matrices. The
simulation uses the dimensionless units $L=C_R=1$, a ring with $M=3$ and
its mode $k=2\pi/3$, for which $\omega_0=\sqrt3$, and the layer duration
$\tau=9.03/\omega_0$. Rescaling the time scale of the circuit as a whole
leaves the results unchanged, so they depend only on $\omega_0\tau$ and not
on the absolute frequency; for example, for $f_0=\omega_0/2\pi=1$~MHz, one
layer lasts about $1.4~\mu$s. Without loss, the growth of the output power
with time agrees with the layer matrices layer by layer
[Fig.~\ref{fig:S11}(a)]. Because the transitions take a finite time, the
growth rate for the length $N=60$ is slightly below the result of the layer
matrices, by at most $0.015$, and the difference tends to zero linearly
with the transition time; within each integer plateau this difference
hardly changes with $h$, so within the plateaus the response differs from
the result of the layer matrices by at most $0.002$ and forms the same
integer plateaus [Figs.~\ref{fig:S11}(b) and~\ref{fig:S11}(c)]. For
$Q=10$, the growth rate drops as a whole by about $0.45$, and the integer
plateaus do not change.

\paragraph{Input and readout.}
Before the modulation starts and after it ends, $a_m=b_m=0$, the two rings
are not coupled, and both are ordinary transmission-line rings. The
forward and backward waves then travel clockwise and counterclockwise
around the rings, and a weakly coupled directional coupler on ring $u$
measures their powers separately; alternatively, the node voltages can be
recorded directly and decomposed afterward into the waves of the two
directions. For a pure forward input, these two powers are
$\mathcal T_N$ and $\mathcal R_N$. A pure forward input requires injecting
clockwise waves into both rings with a quarter-period phase difference.
Injecting into ring $u$ alone also works: the input then contains, in $D$,
the two modes with the wavenumbers $\pm k$, which have the same $\omega_0$
and go through the same layer matrices, so the slope of the growth rate of
the output power with respect to $h$ is still the same integer
(Sec.~S5\,A). The circuit simulation confirms this (caption of
Fig.~\ref{fig:S11}).

\paragraph{Size of the rings.}
The best choice for the number of cells in each ring is $M=3$. The ring
then has, besides the two measured modes $\pm k$, only the mode $k=0$,
whose $\omega_0=0$; this mode neither oscillates nor grows, so no mode
competes with the measured modes. With more cells, the ring has modes of
other wavenumbers; their $\omega_0\tau$ is different, and their growth
rate can be much larger than that of the measured modes. Although the
input does not excite these modes, they grow from noise and after some
layers dominate over the measured modes, so that the measured response is
no longer the integer of the measured modes. Therefore each ring needs
only three cells, and the whole circuit has only six cells.

\paragraph{Experimental requirements.}
The two control waveforms are just $a_m$ and $b_m$ of Eq.~\eqref{eq:S59}
and can be supplied directly by a two-channel arbitrary waveform
generator. We tested the main nonideal factors in the simulation (caption
of Fig.~\ref{fig:S11}). When the gain switching is not fast enough, when
the bandwidth of the controlled sources is limited, or when the gains have
an overall error, the response changes only slightly, and in the MHz range
common analog multipliers meet these requirements. The strictest
requirement is the uniformity of the components among the cells: when the
random deviations reach a few percent, the $n=1$ plateau deviates clearly
from the integer. Each ring has only three cells, however, and selecting
the components to within $1\%$ is not difficult. In addition, $a_m$ and
$b_m$ must be given by Eq.~\eqref{eq:S59} with the same $h$: the integer
response requires that the layer matrix depend on $\varphi$ and $h$ only
through the complex phase $\vartheta$ (Sec.~S4\,A), and this property no
longer holds if the two gains correspond to different values of $h$.
Finally, the dynamic range can be adjusted with loss. Without loss, the
total gain of a few tens of layers far exceeds the dynamic range of the
instruments; a suitable series resistance lowers the net gain to a range
that standard instruments can measure, while the change of the output
power with $h$ remains large enough to measure the integer slope.

\end{document}